\documentclass[fleqn,usenatbib]{mnras}

\usepackage{newtxtext,newtxmath}

\usepackage[T1]{fontenc}
\usepackage{ae,aecompl}
\usepackage{csquotes}

\usepackage{graphicx}	
\usepackage{amsmath}	
\usepackage{amssymb}	
\usepackage{float}
\usepackage{multicol}
\usepackage{bm}
\usepackage{subfigure}

\title[Differential rotation in neutron stars with open and closed magnetic topologies]{Differential rotation in neutron stars with open and closed magnetic topologies}

\author[F. Anzuini, A. Melatos.]{
F. Anzuini,$^{1}$\thanks{E-mail: fanzuini@student.unimelb.edu.au}
A. Melatos$^{1, 2}$
\\
$^{1}$School of Physics, University of Melbourne, Parkville, Victoria 3010, Australia\\
$^{2}$Australian Research Council Centre of Excellence for Gravitational Wave Discovery (OzGrav), University of Melbourne, \\
 \  Parkville, Victoria 3010, Australia
}

\date{Accepted XXX. Received YYY; in original form ZZZ}

\pubyear{2018}

\begin{document}
\label{firstpage}
\pagerange{\pageref{firstpage}--\pageref{lastpage}}
\maketitle

\begin{abstract}
Analytic arguments have been advanced that the degree of differential rotation in a neutron star depends on whether the topology of the internal magnetic field is open or closed.
To test this assertion, the ideal-magnetohydrodynamics solver PLUTO is employed to investigate numerically the flow of an incompressible, viscous fluid threaded by a magnetic field with open and closed topologies in a conducting, differentially rotating, spherical shell. Rigid body corotation with the outer sphere is enforced  on the Alfv\'en time-scale, along magnetic field lines that connect the northern and southern hemispheres of the outer sphere. Along other field lines, however, the behavior is more complicated. For example, an initial point dipole field evolves to produce an approximately closed equatorial flux tube containing at least one predominantly toroidal and approximately closed field line surrounded by a bundle of predominantly toroidal but open field lines. Inside the equatorial flux tube, the field-line-averaged magnetic tension approaches zero, and the fluid rotates differentially, adjusting its angular velocity on the viscous time-scale to match the boundary conditions on the flux tube's toroidal surface. Outside the equatorial flux tube the differential rotation increases, as the magnetic tension averaged along open field lines decreases. 
\end{abstract}




\section{Introduction}

Differential rotation between the rigid crust and the multiple fluid components of a neutron star has been proposed as one possible driver of rotational glitches \citep{vortex, vortex2,vortex3}. Among other factors, the amount of differential rotation is limited by the magnetic field threading the stellar interior. In general, hydromagnetic forces couple the charged components of the multi-fluid interior to the crust, which is decelerated by electromagnetic braking.
Eventually, if the coupling is rapid, one expects the crust and the charged components to approach corotation, although they may rotate differentially with respect to the  superfluid neutrons.
 The global effectiveness of the coupling depends on the magnetic topology. It has been argued that open topologies cause the charged fluid components to corotate with the decelerating crust on the Alfv\'en time-scale, so that all the magnetically coupled parts of the system spin down together \citep{Easson1979, article, Lasky}. In contrast magnetic flux tubes that close inside the star decouple from the regions threaded by open field lines, allowing differential rotation to persist up to long, viscous time-scales in parts of the star.

Numerical and experimental studies of a magnetized, \mbox{conducting} fluid rotating differentially in a spherical Couette configuration have an extensive history. Quantitative results have been obtained concerning the role of the driving shear and shell thickness \citep{Dormy, Nakabayashi_2, Nakabayashi_1}, topology of the applied field \citep{Dormy, Hollerbach_1, Schmitt_1}, conducting boundary conditions \citep{Dormy, Hollerbach_1, Soward}, superrotating shear layers \citep{Dormy_Jault, Nataf_1, Soward}, nonaxisymmetric instabilities \citep{giss, Hollerbach_2, Couette2}, and magnetorotational instabilities \citep{giss}. In all the above studies, the magnetic field induced by the shear flow is small compared to the applied magnetostatic field. This approximation is suitable in various geophysical and laboratory applications.

Previous studies of how the magnetic topology affects differential rotation \textrm{in the neutron star context} are based on analytic calculations \citep{Easson1979, article, Gogl_2}. In this paper we approach the problem numerically for the first time without limiting the \mbox{analysis} to small induced magnetic fields. Using the finite-difference, Godunov-type, ideal-magnetohydrodynamics (ideal-MHD) solver PLUTO \citep{Mignone}, we perform a structured sequence of numerical experiments to measure and compare the angular velocity shear sustained by open and closed magnetic topologies over the short and long term in a differentially rotating spherical shell. The system represents crudely the rigid crust and multi-fluid outer core of a neutron star. We emphasize, however, that the model is highly idealized: its parameters are unrealistic astrophysically due to computational limitations, the boundary conditions do not capture the full complexity of the interaction between different stellar layers, and the single-fluid MHD equations of motion approximate the full, multi-component physics \citep{Glamp}. A similar, idealized model but without a magnetic field has been used in the past successfully to study neutron star turbulence \citep{Peralta_2005, Peralta_2006, Peralta2_2009}, superfluid spherical Couette flow \citep{Peralta_2009}, pulsar glitch statistics \citep{Melatos_2007} and pulsar glitch recovery \citep{Howitt}. 

The paper is organized as follows. In Section \ref{sec:eom} we formulate the problem in an MHD context and follow previous authors in emphasizing the important role played by magnetic-field-aligned coordinates and field-line integrals when interpreting the numerical results to follow \citep{Easson1979, article, Lasky}. Section \ref{sec:Numerical} outlines the numerical method, the initial and boundary conditions, and the dimensionless parameters in the problem. The latter are ordered the same as in a neutron star, although their dynamic range is much smaller due to computational limitations; for example, the magnetic coupling time-scale is kept shorter than the viscous time-scale, as expected in neutron stars. Section \ref{sec:Viscous} validates the code in the unmagnetized case and sets a baseline against which to compare the magnetized flow. Sections \ref{sec:MHD} and \ref{sec:MHD_diff_rot} calculate the magnetized flow in several closed and open \mbox{geometries} as a function of the magnetic field strength and the \mbox{driving} shear. The degree of differential rotation is quantified in each scenario. In Section \ref{sec:flux_tube} we relate the degree of differential rotation to field-line-averaged tension and use the same diagnostic to study the geometric evolution of the magnetized flow.


\section{Equations of motion}
\label{sec:eom}
We model the stellar interior in an idealized fashion as a spherical Couette system, i.e. a differentially rotating, conducting spherical shell containing an incompressible viscous fluid obeying the equations of ideal MHD. The spherical Couette geometry and boundary conditions are transplanted from previous studies of the global flow pattern in a neutron star \citep{Peralta_2005, Peralta_2006,Howitt}.
The simulated system is related but not identical to that considered by \cite{Easson1979}, \cite{article} and \cite{Lasky}. The main difference is that we neglect the presence of a superfluid, neutral component and consider a dynamical magnetic field which evolves nonlinearly, in contrast to previous perturbative treatments \citep{Easson1979,Lasky}.

\subsection{Magnetic coordinates}
\label{sec:mag_coordinates}

Certain key invariants associated with the presence or absence of differential rotation involve integrals of MHD variables along open or closed magnetic field lines, as specified in Section \ref{sec:line_integrals}. The line integrals are easier to calculate and interpret if we switch to magnetic coordinates. The use of magnetic coordinates is widespread in the analysis of complicated magnetic topologies in tokamaks \citep{lif}.

We adopt an orthogonal, curvilinear set of coordinates $\{\psi, \chi, \phi\}$, where $\psi$ is the stream function, $\chi$ is the arc length along the magnetic field line and $\phi$ is the azimuthal angle copied from spherical coordinates. Following \cite{Lasky}, the line element $dl^2$ is diagonal, with

\begin{equation}
dl^2 = h_{\psi}^2(\psi,\chi)d\psi^2 + h_{\chi}^2(\psi,\chi)d\chi^2 + h_{\phi}^2(\psi,\chi)d\phi^2 \ , 
\label{eq:line_element}
\end{equation}
where $ h_{\psi},h_{\chi}, h_{\phi}$ are the metric functions. 
The gradient operator is given by 

\begin{equation}
\nabla = \frac{\hat{\psi}}{h_{\psi}(\psi,\chi)}\partial_{\psi} + \frac{\hat{\chi}}{h_{\chi}(\psi,\chi)}\partial_{\chi} + \frac{\hat{\phi}}{h_{\phi}(\psi,\chi)}\partial_{\phi} \ , 
\end{equation}
where $\hat{\psi}$, $\hat{\chi}$, $\hat{\phi} $ are the unit vectors along the $\psi, \chi$ and $\phi$ directions respectively. The poloidal field becomes

\begin{align}
\vec{B}_p(\psi,\chi)  =& \nabla\psi \times \nabla\phi \\
  = &\hat{\chi}\Big[h_{\psi}(\psi,\chi)h_{\phi}(\psi,\chi)\Big]^{-1} .
\end{align}

The azimuthal component of the momentum equation plays an important role in the physics studied in this paper. In an inertial reference frame and in spherical coordinates $(r,\theta, \phi)$ it reads
\label{sec:spher_coord} 
\begin{equation}
\begin{split}
 \partial_tv_{\phi}& =  \frac{1}{4\pi\rho}(\vec{B}_p \cdot \nabla)B_{\phi} + \mu \nabla^2 v_{\phi} - (\vec{v}\cdot\nabla)v_{\phi} -\hat{\phi} \cdot\nabla p \ ,
 \end{split}
 \label{eq:MHD_spher}
\end{equation}
where $v_{\phi}$ symbolizes the fluid velocity in the azimuthal direction, $\rho$ is the constant and uniform density of the incompressible fluid [suitable for subsonic flow; see \cite{Peralta_2005}], $\vec{B}_p$ and $B_{\phi}$ are the poloidal and toroidal magnetic field components respectively, $\mu$ is the kinematic viscosity coefficient (also assumed to be constant and uniform) and $p$ is the total pressure. As the system is axisymmetric, we have $\hat{\phi}\cdot\bold{\nabla}p = 0$, and (\ref{eq:MHD_spher}) in magnetic coordinates reduces to
\begin{align}\nonumber
\partial_{\chi} B_{\phi}(\psi,\chi)  =&   4\pi \rho h_{\psi}(\psi,\chi)h_{\phi}(\psi,\chi)h_{\chi}(\psi,\chi)\Big[\partial_tv_{\phi}(\psi,\chi) \\
& - \mu \nabla^2 v_{\phi}(\psi,\chi) + \vec{v}(\psi,\chi)\cdot\nabla v_{\phi}(\psi,\chi)\Big] .
\label{eq:MHD_magn}
\end{align}


\subsection{Field line integrals}
\label{sec:line_integrals} 
In magnetic coordinates the left-hand side of \eqref{eq:MHD_magn} reduces to a total derivative. Integrating \eqref{eq:MHD_magn} along an arbitrary \textit{closed} magnetic field line labeled by $\psi$ we get 
\begin{equation}
0 = \oint{d\chi\partial_{\chi} B_{\phi}(\psi,\chi)} \  .
\label{eq:magnetic_loop_integral}
\end{equation}
Physically, the magnetic tension averages to zero when integrated along closed magnetic field lines:

\begin{align}\nonumber
 0  = & \oint{d\chi h_{\psi}(\psi,\chi)h_{\chi}(\psi,\chi)h_{\phi}(\psi,\chi)\Big[\partial_tv_{\phi}(\psi,\chi)}  \\
& - \mu \nabla^2 v_{\phi}(\psi,\chi) + \vec{v}(\psi,\chi)\cdot\nabla v_{\phi}(\psi,\chi)\Big] .
\label{eq:rhs}
\end{align}
\\
As $\psi = $ constant labels an arbitrary field line, equation (\ref{eq:rhs}) holds only if the expression in square brackets vanishes for all $\psi$. 
The vanishing of the bracketed term corresponds to azimuthal momentum conservation in an unmagnetized viscous fluid in a field-line-averaged sense. Hence those regions of the fluid threaded by a magnetic field with a closed topology evolve, as if the magnetic field is absent. The result remains true approximately, if a magnetic field line is nearly closed, as we describe in Section \ref{sec:flux_tube}.

In contrast, an \textit{open} (e.g. dipole) topology leads to

\begin{equation}
0 \neq \int{d\chi\partial_{\chi} B_{\phi}(\psi,\chi)} 
\label{eq:magnetic_open_integral}
\end{equation}
in general. In Sections \ref{sec:MHD} and \ref{sec:MHD_diff_rot} we investigate numerically how an open magnetic field modifies the flow, when the magnetic footpoints terminate on the outer sphere only and when they connect the inner and outer spheres.

\section{Numerical method}
\label{sec:Numerical}

\subsection{PLUTO}
The PLUTO solver \citep{Mignone} integrates a closed set of MHD conservation laws using a finite-volume formalism. Volume-averaged conserved quantities are converted into primitive variables. Flux differences at cell interfaces are computed by solving the associated Riemann problem. We adopt a static grid stretched radially close to the inner and outer boundaries (to resolve viscous boundary layers). The typical grid resolution ranges between $252$ and $300$ points in radius, $206$ points in latitude, and up to $20$ points in azimuth $\phi$. (The initial and boundary conditions enforce axisymmetry. The reader is referred to Section \ref{sec:Nonaxisymmetry} for some preliminary tests of nonaxisymmetric configurations). We select the PLUTO option of a Lax-Friederichs-Riemann solver to compute the flux differences and a second-order Runge-Kutta algorithm to advance the solution in time. The time step is controlled by the Courant-Friederichs-Lewy condition. The typical step is of order $O(10^{-3})$ in dimensionless units, spanning the range $1.4 \times 10^{-3} \lesssim \Delta t \lesssim 3.2 \times 10^{-3}$. The maximum adaptive time-step adjustment is fixed to $1.1$. We verify that the flow is resolved through a sample of higher-resolution tests. \\

\subsection{Initial and boundary conditions}
We start the simulations with $\vec{v} = 0$, such that incompressibility $(\nabla\cdot \vec{v} = 0)$ is satisfied at the outset and henceforth. The two spheres at $r = R_i$ (inner sphere) and $r = R_o$ (outer sphere) rotate with angular velocities $\Omega_i >\Omega_o$ and $\Omega_o$ respectively. We define the rotational shear $\Delta\Omega = (\Omega_i-\Omega_o)/\Omega_i$, and neglect the back reaction of the magnetic and viscous torques, so that $\Delta\Omega$ is fixed.

At the boundaries we impose no-penetration ($v_r = 0, v_{\theta} = 0$) and no-slip ($v_{\phi} =  \Omega_{i,o} R_{i,o} \sin\theta$) conditions for all $t$. We model the fluid and the inner and outer boundaries as perfect conductors. The magnetic field is anchored to the boundaries and frozen into the fluid. We denote with $B_0$ the characteristic magnetic field strength at $t = 0$.

The characteristic length scale of the system is given by $L = R_o- R_i$, and the fluid velocity is scaled by $\Omega_iL$. The Reynolds number is then 

\begin{equation}
Re = \frac{\Omega_iL^2}{\mu}
\end{equation}
and the characteristic viscous time-scale is $t_{\mu} = L^2/\mu$. We scale the magnetic field strength in units of $(4\pi\rho)^{1/2}\Omega_iL$, so that a unit magnetic field corresponds to $v_A = \Omega_iL$, where $v_A$ is the Alfv\'en speed. 

Taken at face value, neutron star interiors have high Reynolds numbers $(Re \approx 10^{11})$ \citep{Rey2, Melatos_2007}. It is possible to argue to the contrary, that $Re$ is low in the frame of the stellar crust, given a shear of $10^{-10}\leq\Delta\Omega/\Omega\leq10^{-5}$ as inferred from measurements of rotational glitches \citep{Espinoza_1}. However, this assumption is debatable, as glitches may relax only a small fraction of the shear, and the poorly known effective viscosity of the neutron superfluid component may be lower than expected. Either way, computational limitations restrict us to $Re =  500$ in order to avoid unresolvable turbulent eddies and non-axisymmetric instabilities \citep{Nakabayashi_3, Skinner_1, giss}. For the same reasons, the shell cannot be too thick \citep{ Nakabayashi_2,Nakabayashi_1}. We take $\delta = (R_o - R_i)/R_i = 1.86$ in this paper \citep{Peralta_2005}. A brief discussion of how higher Reynolds numbers (or a tilted magnetic axis) lead to hard-to-resolve \mbox{nonaxisymmetric} and even turbulent flows is presented in Section \ref{sec:Nonaxisymmetry}.

\subsection{Magnetic and viscous time-scales}

Alfv\'en waves are expected to cross a neutron star on time-scales $t_A \leq 10$ s. The high Reynolds number implies that the viscous time-scale $t_{\mu}$ is many orders of magnitude longer than $t_A$ \citep{Easson1979, article, Lasky}. In the presence of differential rotation, there is also an intermediate time-scale, the Ekman time $t_E = t_{\mu}Re^{-1/2} \approx \Omega_i^{-1}Re^{1/2}$, which lies between $t_A$ and $t_\mu$. In our simulations we respect the above ordering, i.e. $\Omega_i^{-1} \lesssim t_A \lesssim t_E \lesssim t_{\mu}$, except for weakly magnetized validation and calibration runs with $B_0 \lesssim 10^{-3}$. The ratio between $t_A$ and $t_\mu$ reads

\begin{equation}
\frac{t_A}{t_{\mu}} = \Big(B_0 Re\Big)^{-1} \ ,
\end{equation}
where $B_0$ is measured in units of $(4\pi\rho)^{1/2}\Omega_iL$ as above.

\begin{figure}
\begin{center}
\includegraphics[width=8cm, height = 7 cm]{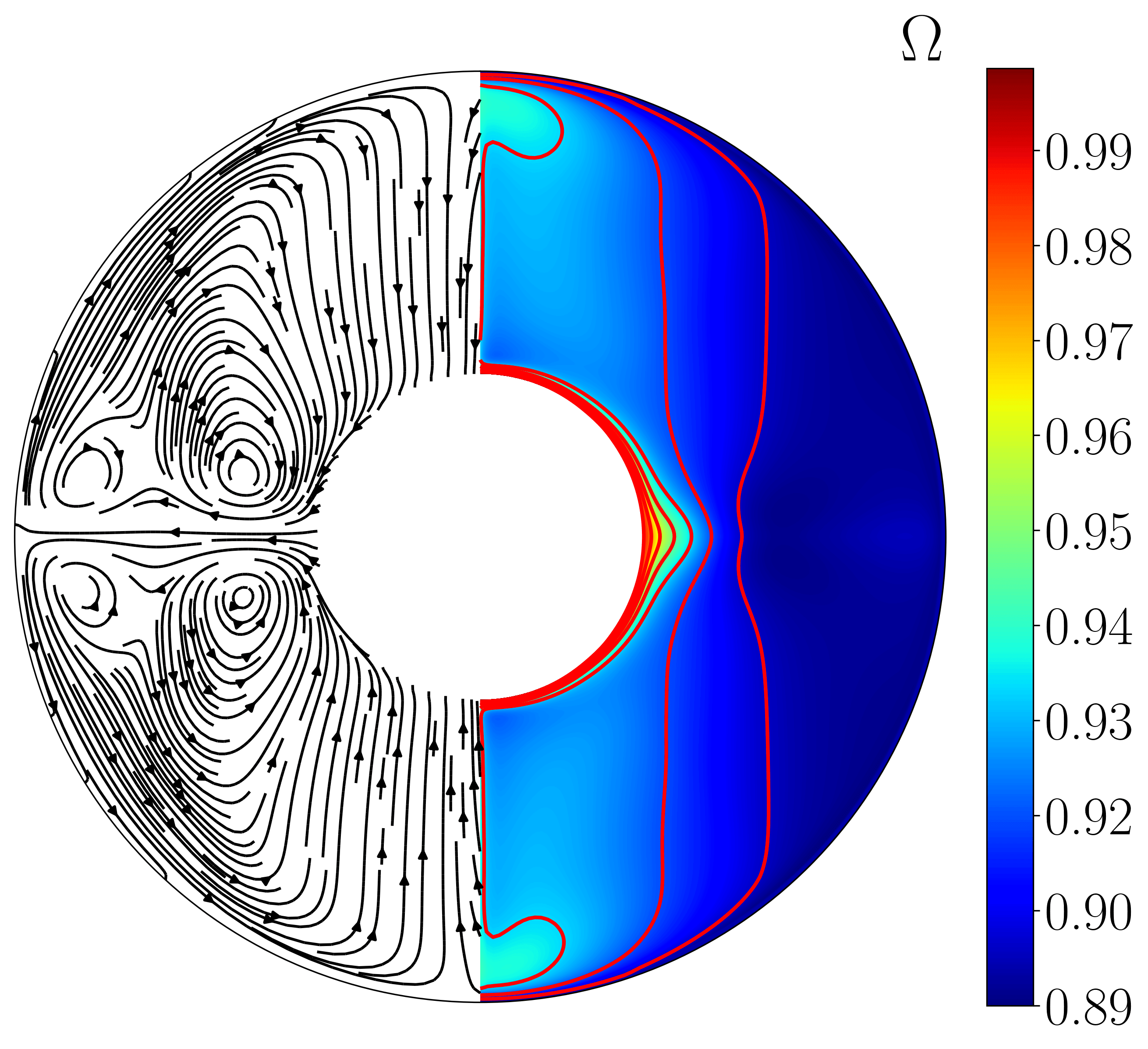}
\caption{Primary rotation and secondary flow driven by Ekman pumping for an unmagnetized, viscous fluid in spherical Couette geometry with $\Delta\Omega = 0.1$ at $t = 1.4 t_{\mu}$. (\textit{Right half}) Contours of the angular velocity $\Omega$ (color bar in units of $\Omega_i$). The red curves mark the contour levels in steps of $0.01 \Omega_i$. (\textit{Left half}) In-plane streamlines. The fluid satisfies the Taylor-Proudman theorem; the angular velocity is almost independent of the latitude in much of the shell. The figure looks identical for $B_0 = 0$ and $1.5\times10^{-5}\leq B_0\leq 1.5\times10^{-2}$, if $\vec{B}$ is purely toroidal at $t = 0$. }
\label{fig:UN_closed_delta01}
\end{center}
\end{figure}

\section{VALIDATION: UNMAGNETIZED FLOW}
\label{sec:Viscous}

We test the numerical code by comparing its output with unmagnetized pseudospectral simulations obeying the same initial and boundary conditions \citep{Peralta_2005, Peralta_2009, Howitt}. As well as validating PLUTO for our problem against an independent solver, this also provides a control experiment which sets a baseline against which we compare the magnetized flow.

In spherical Couette flow, in the regime of fast rotation and low viscosity, the Taylor-Proudman theorem \citep{proudman} states that the flow is approximately azimuthal and columnar (i.e. gradients are small parallel to the rotation axis). The imposed differential rotation drives a secondary flow via Ekman pumping, with $v_r/v_{\phi} \propto Re^{-1/2}$. Figure \ref{fig:UN_closed_delta01} confirms this behavior for $\Delta\Omega = 0.1$. The right half of the meridional plane displays angular velocity contours, which are approximately columnar. A cylindrical Stewartson layer touches the inner sphere at the equator. Fluid rotates with $v_{\phi} \approx \Omega_iR_i$ inside the Stewartson layer and $v_{\phi} \approx \Omega_oR_o$ outside the Stewartson layer. The left half of the meridional plane displays streamlines of the meridional recirculation. The fluid is pumped from the inner, faster sphere down towards the equator, then outwards to the outer, slower sphere, and thence to the poles. Close to the equator, the streamlines trace two circulation cells per hemisphere, where the fluid is sucked out of the equatorial plane, driven to higher latitudes, and recycled back to the equator. 

Next we verify that the flow in a purely toroidal magnetic field evolves, as if the magnetic field is absent. 
To do so, we produce a version of Figure \ref{fig:UN_closed_delta01} with $\Delta\Omega = 0.1$, $1.5\times10^{-5}\leq B_0\leq 1.5\times10^{-2}$, and $\vec{B}_p = 0$ at $t = 0$. The result (not shown) is completely indistinguishable from Figure \ref{fig:UN_closed_delta01}.
This is because (i) the magnetic tension term in equation (\ref{eq:MHD_spher}) vanishes locally for $\vec{B} = B_{\phi}\hat{\phi}$, independent of $B_0$, and (ii) the form of the induction equation guarantees $\vec{B}_p = 0$ at $t>0$ given $\vec{B}_p = 0 $ at $t = 0$.

\begin{figure}
\begin{center}
\includegraphics[width=8.5 cm, height = 8.0 cm]{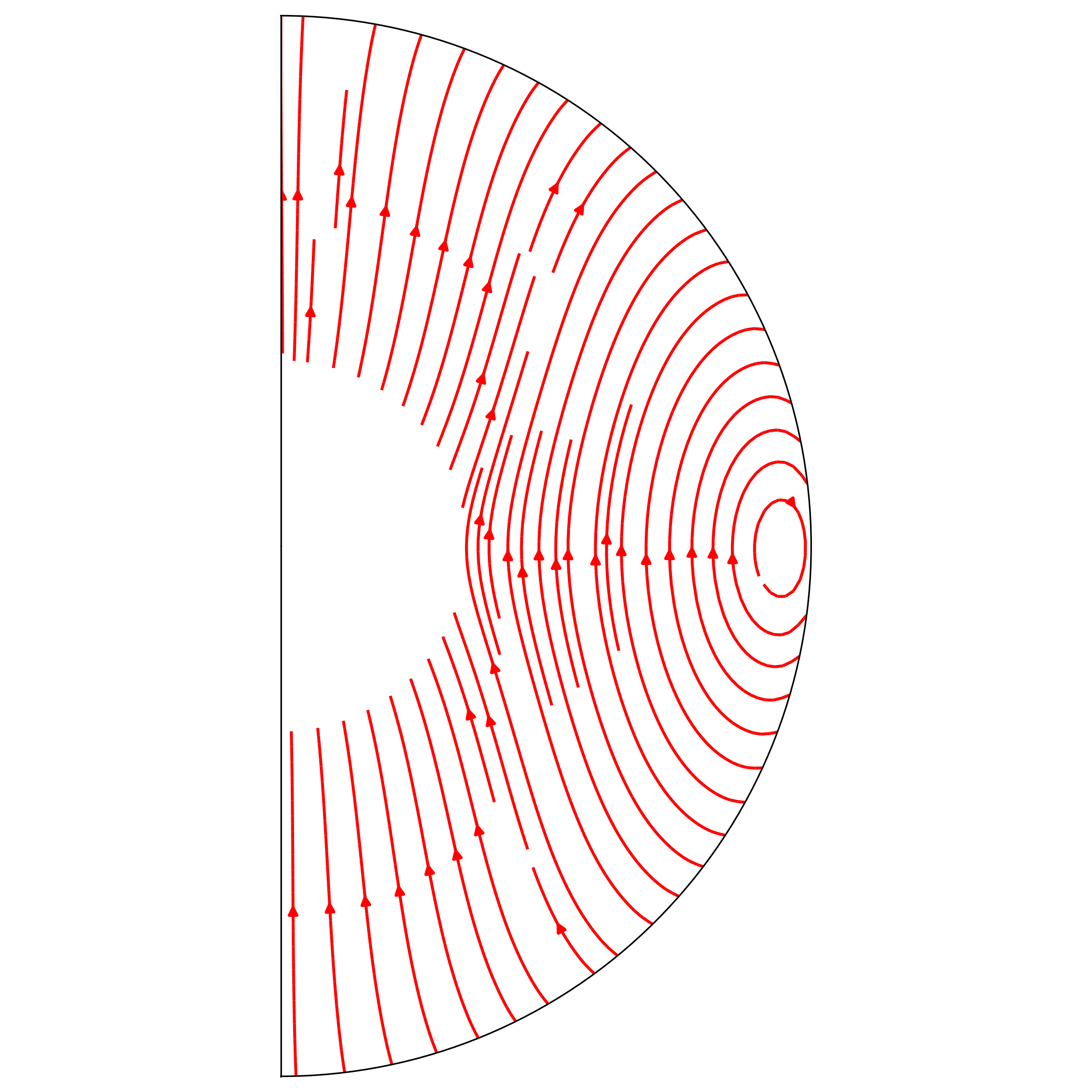}
\caption{Magnetic topology of the initial dipole configuration given by (\ref{eq:dipole}). Near the poles, open field lines connect the differentially rotating inner and outer spheres. At lower latitudes, open field lines connect the northern and southern hemispheres of the outer sphere without touching the inner sphere. Some small, closed poloidal loops also exist at $r \approx R_o, \theta \approx \pi/2$. }
\label{fig:initial_dipole}
\end{center}
\end{figure}

\section{MAGNETIZED FLOW: COROTATION WITH THE CRUST}
\label{sec:MHD}

\begin{figure*}
\begin{center}
\includegraphics[width=16cm, height = 12 cm]{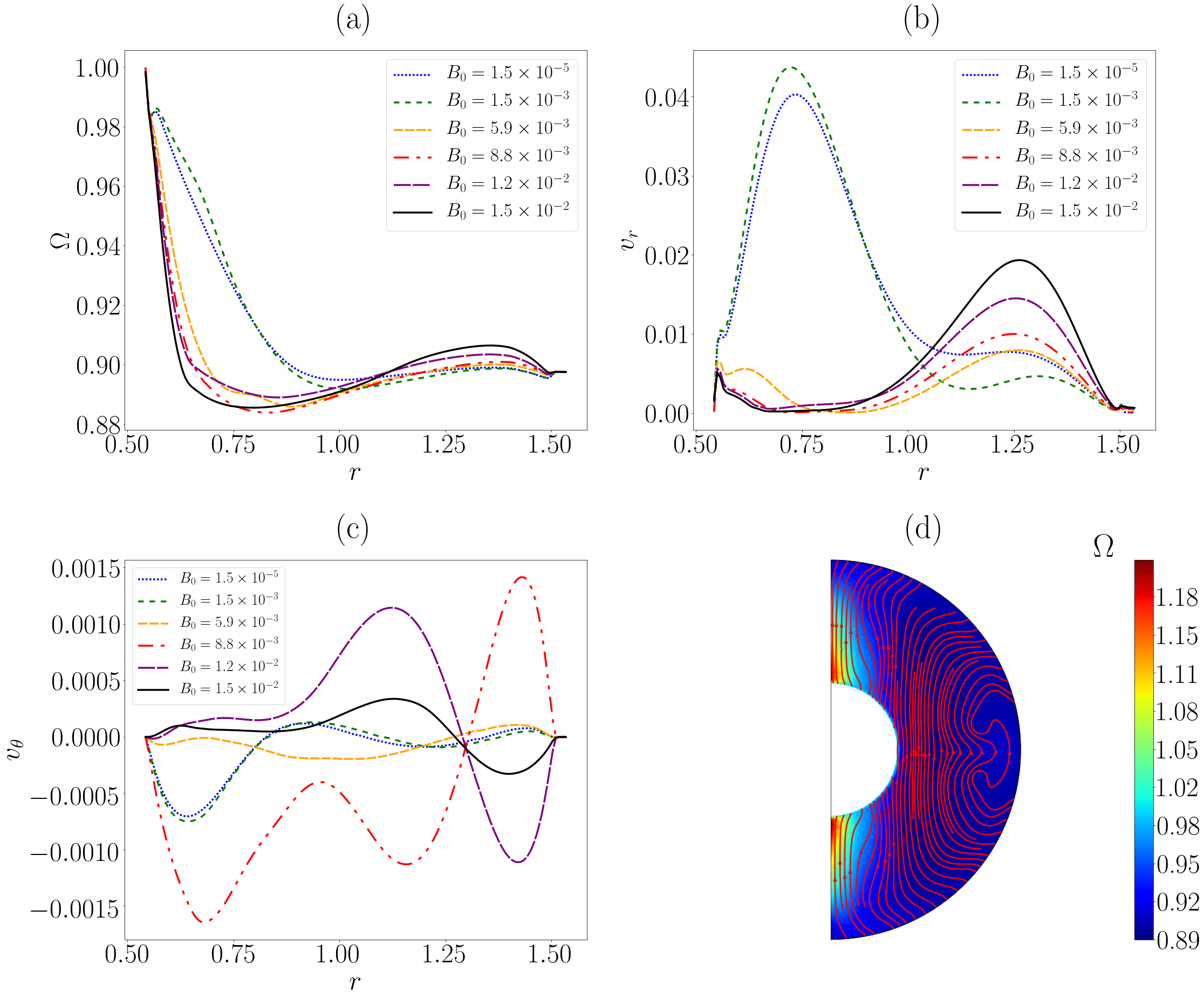}
\caption{Approximate corotation with the outer sphere for an initial dipole magnetic field (initial strength $1.5\times10^{-5} \leq B_0 \leq 1.5\times10^{-2}$). (a) Angular velocity as a function of radius $r$ at the equator ($\theta \approx \pi/2$). As $B_0$ increases, more fluid corotates with the outer shell. (b) Radial velocity component as a function of $r$ at $\theta \approx \pi/2$. (c) Latitudinal velocity component as a function of $r$ at $\theta \approx \pi/2$. (d) Angular velocity contours for  $B_0 = 1.5\times10^{-2}$, with poloidal magnetic field lines overplotted (red contours). The closed poloidal loops at $\theta \approx \pi/2$ are pushed closer to the outer boundary by the open field lines with endpoints at $\theta \approx 5\pi/12$ and $\theta \approx 7\pi/12$ on the outer sphere. All panels are snapshots at $t = 1.4 t_{\mu}$ for $\Delta\Omega = 0.1$. }
\label{fig:initial_dipole_all_delta01}
\end{center}
\end{figure*}

We now turn on the magnetic field and explore numerically the effect of its topology on the flow. 

In this section, we study a topology that is predominantly open initially and remains so throughout the simulation. We initialise the simulation with a field of the form 

\begin{equation}
\vec{B} = \frac{B_0}{2}\Bigg\{\Big[5 - \Big(\frac{r}{R_o}\Big)^2\Big]\cos\theta, \Big[6\Big(\frac{r}{R_o}\Big)^2-5\Big]\sin\theta, 0\Bigg\} \ ,
\label{eq:dipole}
\end{equation}
whose latitudinal variation is dipolar. The configuration \eqref{eq:dipole} is plotted in Figure \ref{fig:initial_dipole}. 
Throughout most of the volume, the field lines do not touch the inner sphere; they connect the outer northern hemisphere to the outer southern hemisphere. Near the poles, the field lines do connect the inner and outher spheres and are also open. Finally, there are also some small, closed poloidal loops straddling the equatorial plane near the outer sphere, which occupy a small fraction of the total volume.

Magnetic tension enforces corotation of the fluid with the outer sphere, along those field lines that do not touch the inner sphere.
Figure \ref{fig:initial_dipole_all_delta01} displays the angular velocity $\Omega$, and linear velocity components $v_r$ and $v_{\theta}$ on the equatorial plane, as functions of $r$. 
 When the magnetic field is weak, i.e. $B_0 \leq 1.5\times 10^{-3}$, the solution cannot be distinguished from the unmagnetized one (blue curve) or else resembles it closely (green curve). However, for $B_0 \geq 5.9\times 10^{-3}$, there are clear differences. Figure \ref{fig:initial_dipole_all_delta01} (a), which plots $\Omega(r)$, shows that $\vec{B}$ enforces corotation with the outer sphere over a wider range of radii ($0.65 \lesssim r \lesssim 1.54$) than in the unmagnetized flow ($1.0 \lesssim r \lesssim 1.54$).

 Figure \ref{fig:initial_dipole_all_delta01} (b) shows the corresponding behavior for $v_r$. In the weak field regime $B_0 \leq 1.5\times10^{-3}$ (blue and green curves), $v_r$ peaks due to Ekman pumping at the Stewartson layer near the inner sphere ($r \approx 0.75$) as in the unmagnetized flow in Figure \ref{fig:UN_closed_delta01}. For $B_0 \geq 5.9\times 10^{-3}$, however, $v_r$ is suppressed near the inner sphere and peaks near the outer sphere ($r \approx 1.3$). The peak shifts because the secondary circulation cells move away from the inner sphere towards the outer sphere. The maximum $|v_r|$ in this radial jet increases with $B_0$. The $v_{\theta}$ component in Figure \ref{fig:initial_dipole_all_delta01} (c) is between one and two orders of magnitude smaller than $v_r$. The equatorial symmetry of the system imposes $v_{\theta} = 0$ at the equator, separating the circulation in the two hemispheres. For $B_0 \leq 1.5\times 10^{-3}$, $v_{\theta}$ closely follows the unmagnetized solution. The peak in $v_{\theta}$ does not move monotonically with $B_0$. Increasing $B_0$ from $5.9\times10^{-3}$ up to $1.2\times10^{-2}$ produces local extrema of $v_{\theta}$ at $r \approx 1.1$ and $r \approx 1.4$, which are suppressed when $B_0$ increases further to $1.5\times10^{-2}$.
  
Figure \ref{fig:initial_dipole_all_delta01} (d) demonstrates how $\Omega(r,\theta)$ (color scale) relates to the poloidal magnetic field lines (red contours). It generalizes Figure \ref{fig:initial_dipole_all_delta01} (a) to show  $\Omega(r,\theta)$ throughout the spherical shell for $B_0 = 1.5 \times 10^{-2}$ initially and $\Delta\Omega = 0.1$, i.e. when the magnetic field is dynamically important. In Appendix \ref{sec:all_dipole}, for completeness, we present analogous plots for lower magnetic fields in the range $1.5 \times 10^{-5}\leq B_0 \leq 1.5 \times 10^{-2}$. The dark blue region, in which the fluid corotates with the outer boundary, is threaded by magnetic field lines that connect the northern and southern outer hemispheres without touching the inner shell. Magnetic tension enforces approximate corotation with the outer shell. Open field lines that touch the outer shell at latitudes $\theta \approx 5\pi/12$ and $\theta \approx 7\pi/12$ are deformed close to the equator, squeezing the poloidal loop on the outer boundary. Accordingly, $\Omega$  rises slightly at $r \approx 1.3$ as seen in Figure \ref{fig:initial_dipole_all_delta01} (a). The angular velocity exceeds $\Omega_i >\Omega_o$ by up to $20\%$ close to the poles, where magnetic field lines connect the inner and the outer sphere.

The reader may wonder why $v_{\theta}$ does not vanish at $\theta = \pi/2$ in Figure \ref{fig:initial_dipole_all_delta01}(c), as expected from north-south symmetry. Do we introduce a symmetry-breaking perturbation into the system (beyond those that arise unavoidably from numerical errors)? The answer is no: $v_{\theta}$ is not \mbox{exactly} zero because it is sampled slightly away from the equator, in a region where the gradient $\partial v_{\theta}/\partial{\theta}$ is high. The curves in Figure \ref{fig:initial_dipole_all_delta01}(c) are for $\theta = 1.5733$ rad $\neq \pi/2$, which is as close as one gets to the equator with the grid placement and resolution chosen ($206$ points in $\theta$). By interpolating the solution across the equator, we obtain $v_{\theta} \approx 0$ to a good approximation at $\theta = 1.5708 = \pi/2$ \mbox{exactly}, with $|v_{\theta}(\theta = \pi/2)|\leq 0.01|v_{\theta}(\theta = 1.5733)|$. Furthermore, the $v_{\theta}$ pattern is north-south symmetric but switches in sign about the equator, as expected for equatorial symmetry. Secondary flows circulate fluid from the equator towards the north pole in the northern hemisphere, and from the equator to the south pole in the southern hemisphere, as in unmagnetized spherical Couette flow.

\section{MAGNETIZED FLOW: DIFFERENTIALLY ROTATING, MAGNETICALLY COUPLED CRUST AND CORE}
\label{sec:MHD_diff_rot}

\begin{figure}
\begin{center}
\includegraphics[width=8.5 cm, height = 8.0 cm]{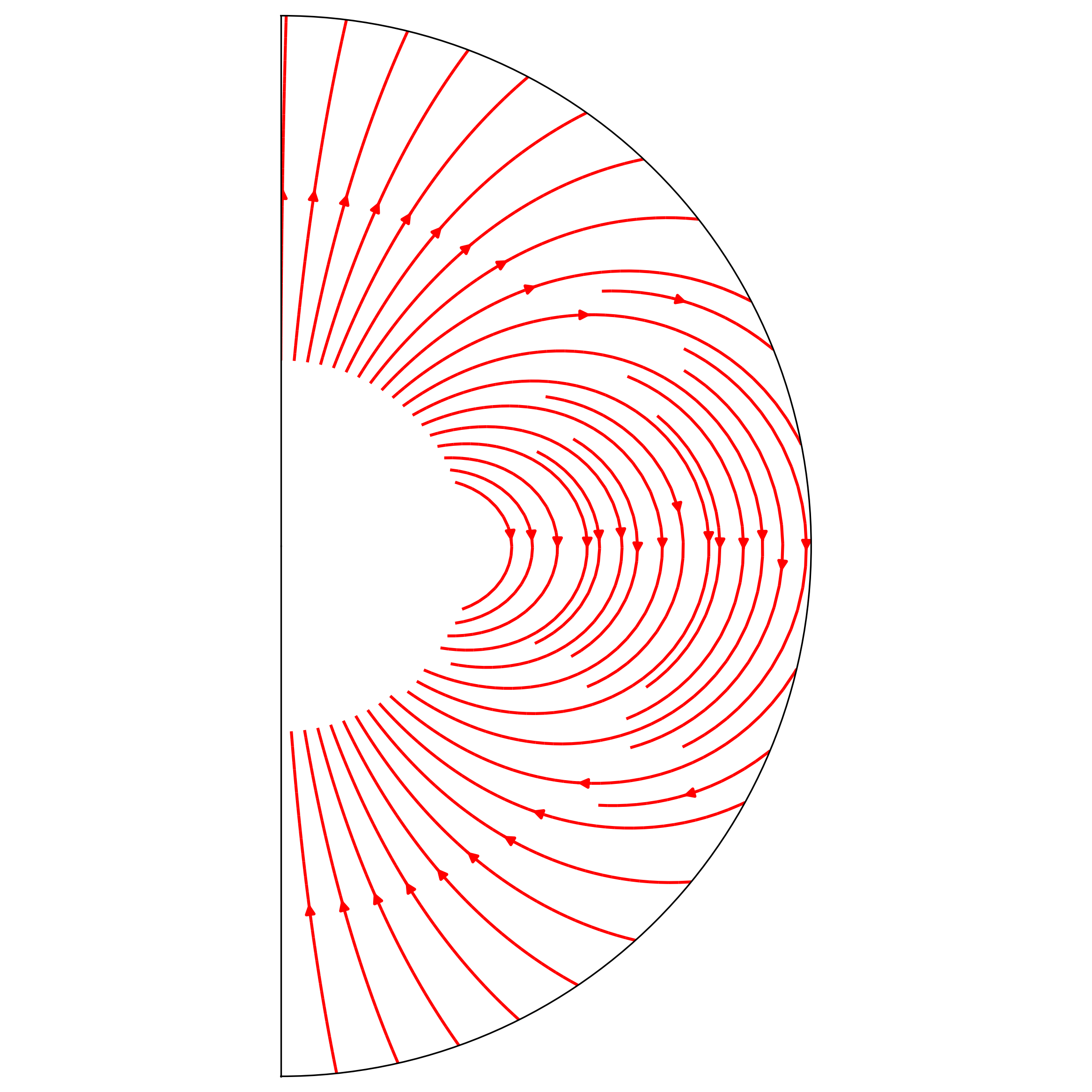}
\caption{Magnetic topology of the initial dipole configuration given by (\ref{eq:dipole_internal_sources}). Near the poles, open field lines connect the inner and outer spheres. At lower latitudes, the magnetic field lines connect the inner sphere to itself. }
\label{fig:central_point_dipole}
\end{center}
\end{figure}

\begin{figure}
\begin{center}
\includegraphics[width=8.5 cm, height = 20 cm]{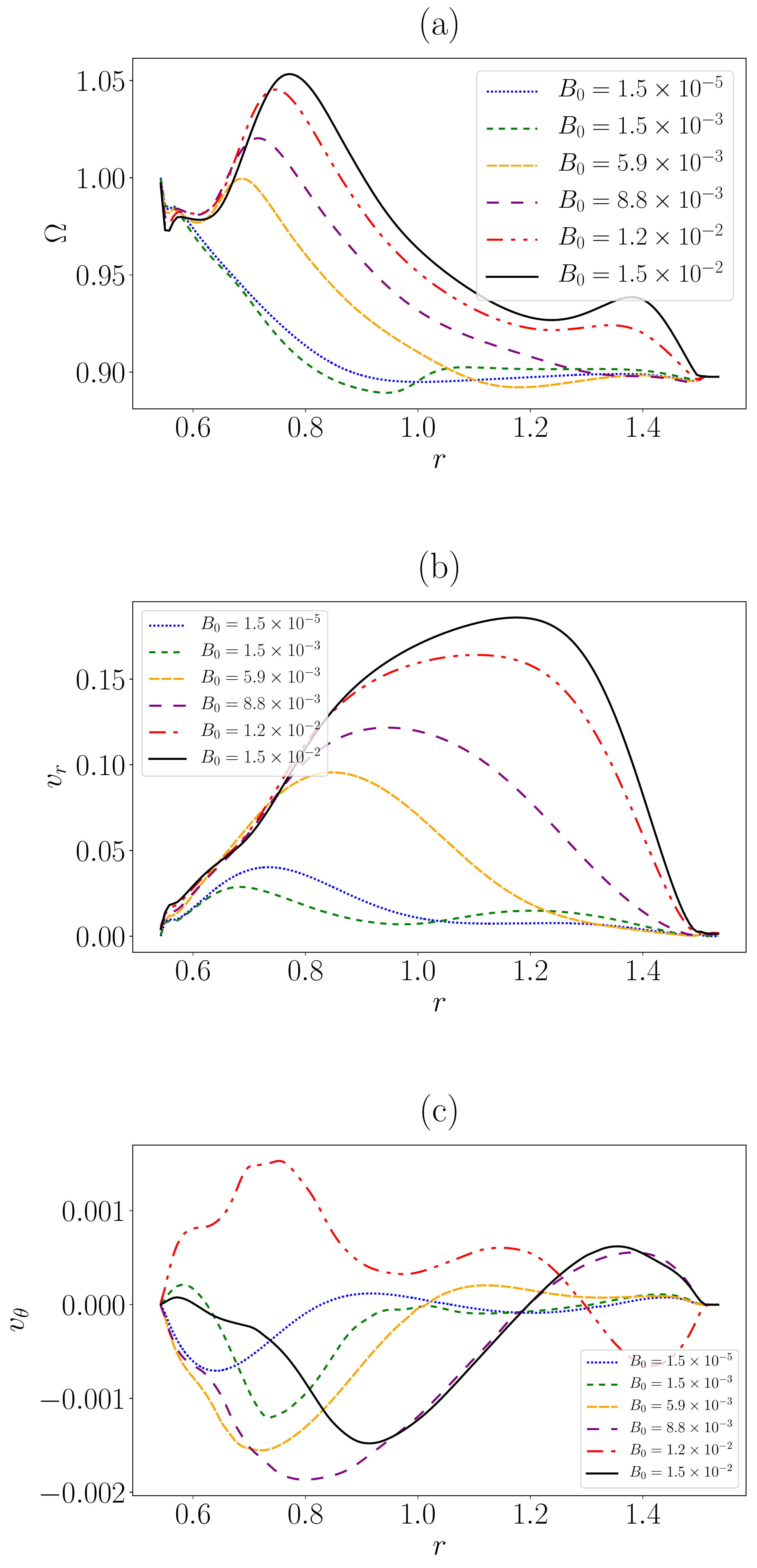}
\caption{(a) Angular velocity as a function of $r$ at the equator ($\theta \approx \pi/2$). As the magnetic field strengthens, one finds max($\Omega$) $ > \Omega_i$.  (b) Radial velocity of the fluid as a function of the radius at $\theta \approx \pi/2$. (c) Latitudinal velocity $v_{\theta}(r)$ for $\theta \approx \pi/2$. $B_0$ varies in the range $[1.5\times10^{-5}, 1.5\times 10^{-2}]$.  The plots are snapshots taken at $t = 1.4 t_{\mu}$ for $\Delta\Omega = 0.1$. }
\label{fig:vr_omega_vtheta_delta01}
\end{center}
\end{figure}

We switch now to an initial magnetic field of the dipolar form \citep{Nataf_1, Schmitt_1, Dormy, Dormy_Jault, Hollerbach_1, Soward}

\begin{equation}
\vec{B} = \frac{B_0}{2r^3}\Big(2\cos\theta, \sin\theta, 0\Big) \ ,
\label{eq:dipole_internal_sources}
\end{equation}
drawn in Figure \ref{fig:central_point_dipole}. Physically (\ref{eq:dipole_internal_sources}) corresponds to a point dipole, whose source currents are localized at $r = 0$. In (\ref{eq:dipole}), open field lines either connect the inner and outer spheres or the northern and southern hemispheres of the outer sphere. In \eqref{eq:dipole_internal_sources} open field lines at high latitudes join the inner and outer spheres, whereas closed field lines near the equator connect the northern and the southern hemispheres of the inner sphere. By imagining that these field lines extend inwards to connect the two footpoints within the inner sphere, we see that field lines in this category effectively form closed loops. It would be preferable to double check the results for open and closed field lines lying entirely within the simulation volume. However, this test is impossible, because it is a fundamental property of spherical Couette flow that it is numerically unstable for $R_i\rightarrow 0$ (indeed $R_i < 0.5 R_o$); see the reviews by \cite{Nakabayashi_2} and \cite{Nakabayashi_1}. As in previous papers, a spherical Couette arrangement with $\Delta\Omega \neq 0$ is used to approximate a filled sphere ($R_i = 0$) decelerating at a rate $\dot{\Omega}$ with $\Delta\Omega \approx \dot{\Omega}\tau$, where $\tau$ is the characteristic time-scale for corotation to be enforced in the absence of deceleration.\par
Figure \ref{fig:vr_omega_vtheta_delta01} displays $\Omega, v_r$ and $v_{\theta}$ as functions of radius in the equatorial plane for $1.5\times 10^{-5}\leq B_0 \leq 1.5\times 10^{-2} $ and $\Delta\Omega = 0.1$. The physics is similar to Figure \ref{fig:initial_dipole_all_delta01}, but the results are different, because the magnetic topology is different. For $B_0 \leq 1.5\times10^{-3}$, when the magnetic field is unimportant dynamically, the flow closely resembles the unmagnetized case. For $5.9\times 10^{-3}\leq B_0 \leq 1.5\times10^{-2}$, the closed equatorial zone strives to corotate with the inner sphere, in marked contrast to Figure \ref{fig:initial_dipole_all_delta01}. In doing so, it overshoots, and $\Omega$ peaks at $\Omega \approx 1.05$ in the interval $0.7 \lesssim r \lesssim 0.8$. This occurs because currents are free to flow through the rigid, conducting walls of the spherical shell \citep{Soward, Dormy_Jault}. In order to compensate, the fluid generates additional currents via rotation, allowing the fluid to deviate from Ferraro's law of isorotation \citep{Ferraro} and to attain $\max(\Omega) > \Omega_i$ \citep{Nataf_1}. The peak of $\Omega$ moves from $r\approx 0.7$ at $B_0 = 5.9 \times 10^{-3}$ to $r \approx 0.8$ at $B_0 = 1.5 \times 10^{-2}$, as the magnetic coupling to the inner sphere strengthens. As in Figure \ref{fig:initial_dipole_all_delta01} (b), the peak of $v_r \approx 0.1 r\Omega$ at $B_0 = 5.9\times10^{-3}$ shifts outwards, as $B_0$ increases, because the principal circulation cells of the meridional flow move closer to the outer boundary. The $v_{\theta}$ component in Figure \ref{fig:vr_omega_vtheta_delta01} (c) is small, with $|v_{\theta}| \approx 10^{-3} r\Omega$, again because of equatorial symmetry.

\begin{figure}
\begin{center}
\includegraphics[width=8.5 cm, height = 20 cm]{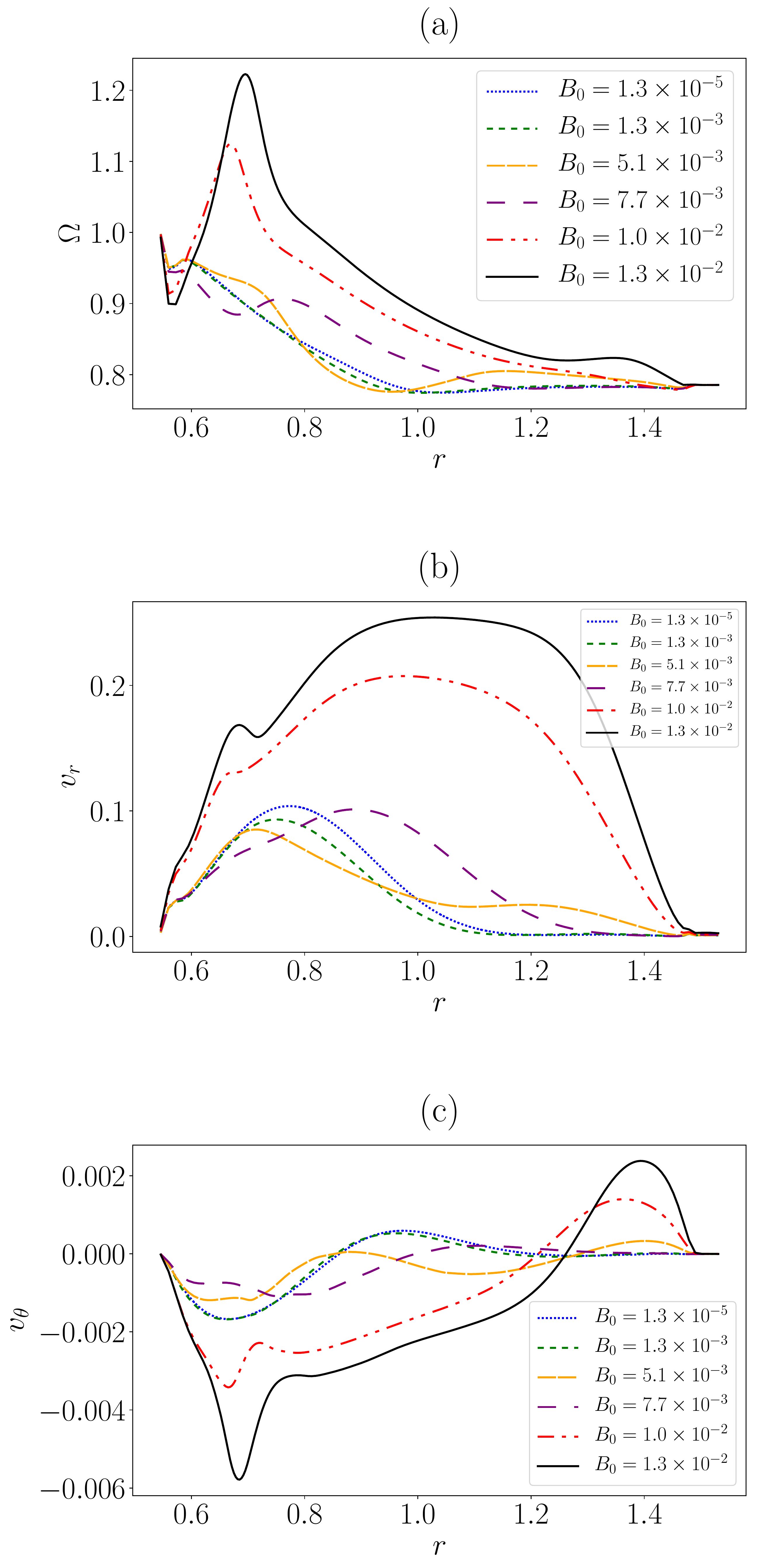}
\caption{Same as in Figure \ref{fig:vr_omega_vtheta_delta01} but with higher rotational shear ($\Delta\Omega = 0.2$) and $t = 1.6 t_{\mu}$.}
\label{fig:vr_omega_vtheta_delta02}
\end{center}
\end{figure}

We now double the shear to $\Delta\Omega = 0.2 $ in Figure \ref{fig:vr_omega_vtheta_delta02}. The greater the shear, the more magnetic field lines joining the differentially rotating shells are stretched, and the harder it is for the magnetic tension to enforce corotation.  
As a result, we find $\Omega > \Omega_i$ for a narrower range of $r$ [$0.6 \lesssim r \lesssim 0.8$; cf. Figure \ref{fig:vr_omega_vtheta_delta01} (a)] and higher values of $B_0$ [$B_0 > 7.7 \times 10^{-3}$; cf. Figure \ref{fig:vr_omega_vtheta_delta01} (a)]. Interestingly, the peak of $\Omega$ is not only narrower but taller, reaching $\max(\Omega) \approx 1.2$ [cf. $\max(\Omega) \approx 1.05$ in Figure \ref{fig:vr_omega_vtheta_delta01} (a)].
Something similar happens to the radial jet in Figure \ref{fig:vr_omega_vtheta_delta02} (b). Its profile exhibits a broad plateau at $0.9 \lesssim r \lesssim 1.2$ for $B_0 =1.3\times10^{-2}$, with $\max (v_{r}) \approx 0.25$, which is broader and faster than the radial jet in Figure \ref{fig:vr_omega_vtheta_delta01} (b). The latitudinal velocity $v_{\theta}$ remains like in Figure \ref{fig:vr_omega_vtheta_delta01} (b) due to equatorial symmetry.
Contours $\Omega(r,\theta)$ for different $B_0$ with $\Delta\Omega = 0.1$ and $\Delta\Omega = 0.2$ are presented in Appendix \ref{sec:all_dipole} for completeness.

The astrophysical interpretation of the above results is as follows. If the inner sphere corresponds to a distinct, rigid, stellar component, such as a crystalline color superconducting core \citep{Alford_1, Alford_2, Mannarelli}, then Figures \ref{fig:vr_omega_vtheta_delta01} and \ref{fig:vr_omega_vtheta_delta02} imply that there is a region of differential rotation at $r \approx R_i$. The differential rotation persists, because the spin-down torque on the outer sphere maintains $\Omega_i \neq \Omega_o$. The magnitude of the lag, $\Omega_i - \Omega_o$, depends on the moment of inertia of the inner sphere, e.g. the size and composition of the crystalline superconducting core. On the other hand, if the inner core is not rigid, $r = R_i$ corresponds to a mathematical surface introduced to promote numerical stability in the simulation \citep{Nakabayashi_1,Nakabayashi_2,Peralta_2005}, and the region $r \leq R_i$ in the real star contains the same fluid as $R_i \leq r \leq R_o$. In the latter scenario, the differential rotation is expected to be weaker than in Figures \ref{fig:vr_omega_vtheta_delta01} and \ref{fig:vr_omega_vtheta_delta02}, as the field lines connecting the outer and inner spheres enforce approximate corotation in the stellar volumes $0 \leq r \leq R_i$ and $R_i \leq r \leq R_o$. However, exact corotation is never achieved for two reasons. First, the hydromagnetic torques do not act instantaneously; in the time $t_A$ that Alfv\'en waves take to propagate into the core, the outer sphere spins down electromagnetically, and one has $\Omega_i - \Omega_o \approx t_A\dot{\Omega}_o$. Second, taking the magnetic topology in Figure \ref{fig:central_point_dipole} as an example, there is a region near the equator that is magnetically disconnected from the outer sphere. This is the region threaded by magnetic field lines that start and end at $r = R_i$ in the (artificial) simulation and close within the core without touching $r = R_o$ in the real star. Corotation in a magnetically disconnected volume takes longer to achieve, as discussed in Section \ref{sec:flux_tube}.

\section{EVOLUTION OF THE MAGNETIC GEOMETRY}
\label{sec:flux_tube}
\subsection{Approximately closed equatorial flux tubes}
\label{sec:Generating_flux_tube}

\begin{figure*}
\begin{center}
\includegraphics[width=14cm, height = 8.5 cm]{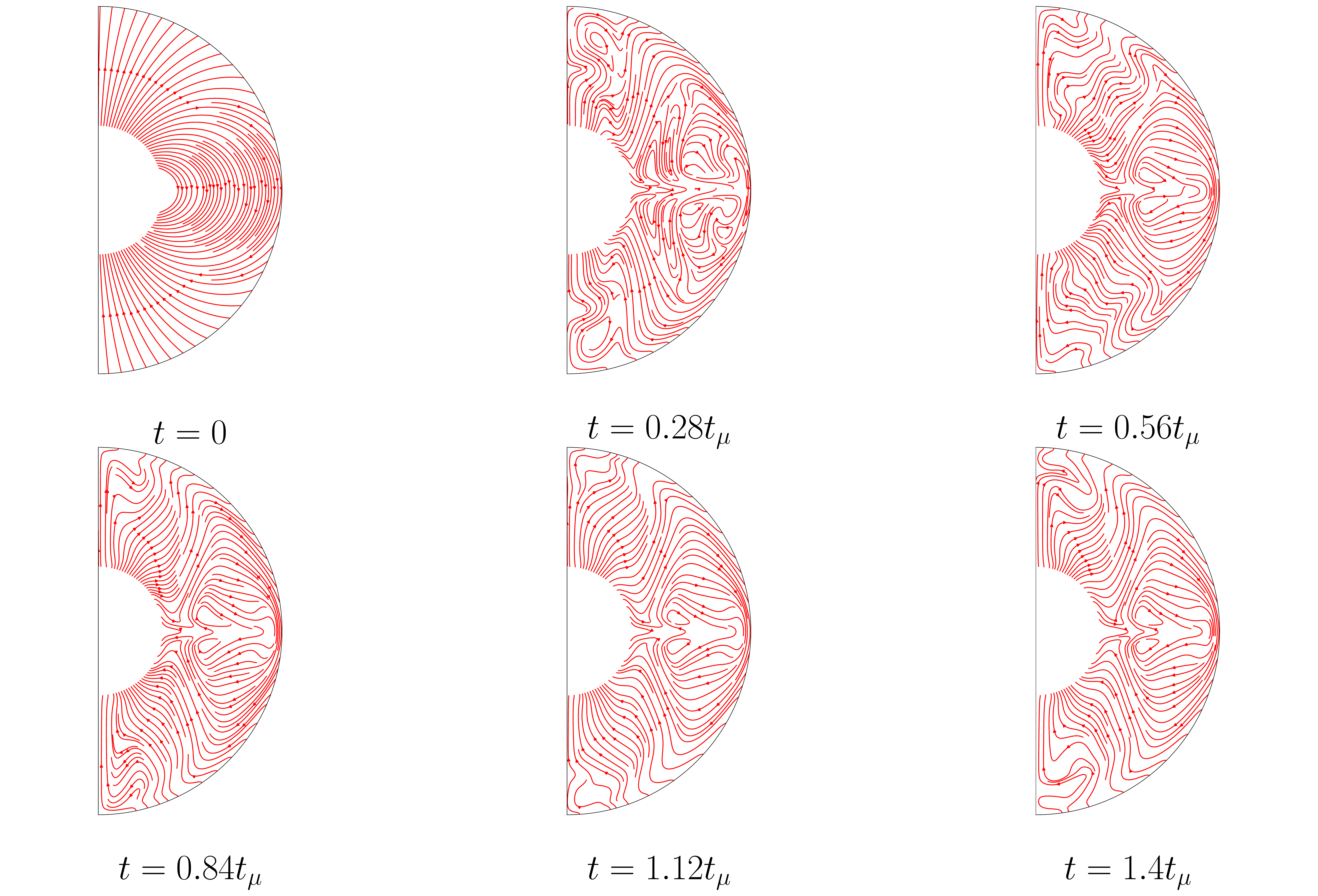}
\caption{Evolution of the magnetic geometry for $\Delta\Omega = 0.1$ and $B_0 = 1.5\times10^{-2}$. }
\label{fig:Topology_Delta01_tfin}
\end{center}
\end{figure*}

\begin{figure*}
\begin{center}
\includegraphics[width=14cm, height = 8.5 cm]{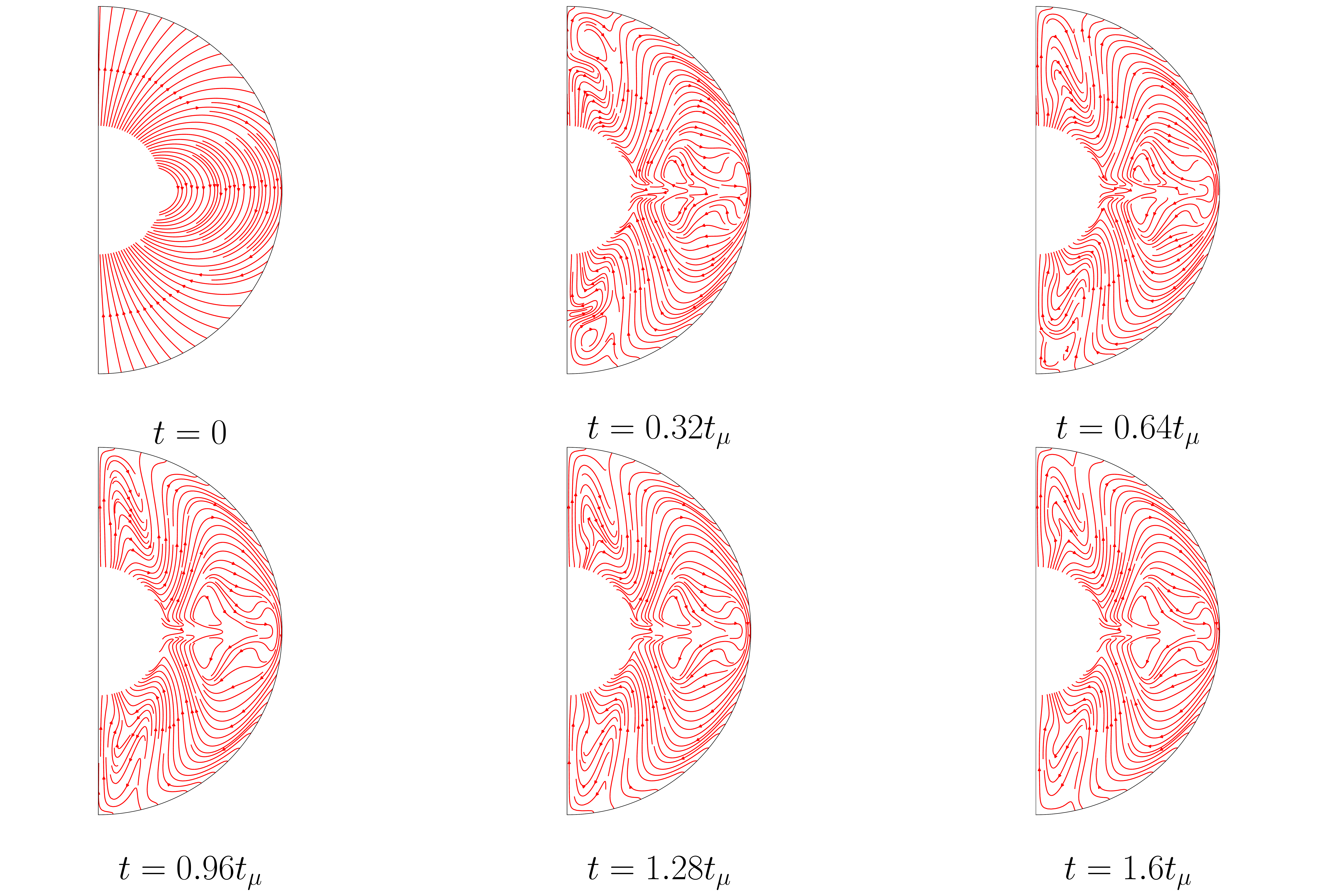}
\caption{Evolution of the magnetic geometry for $\Delta\Omega = 0.2$ and $B_0 = 1.3\times10^{-2}$.}
\label{fig:Topology_Delta02_tfin}
\end{center}
\end{figure*}

Does an initially open magnetic geometry develop closed (or approximately closed) flux tubes under differential rotation? This question is important, because once a flux tube closes, the fluid inside can rotate differentially on the long, viscous time-scale $t_{\mu}$, even if the rest of the star corotates \citep{article,Lasky}. 
 Indeed we show below that the fluid inside a closed flux tube evolves, as if one has $|\vec{B}| \approx 0$ in the tube; it adjusts to the boundary conditions on the flux tube surface (set by the rest of the flow) on the time-scale $t_{\mu}$. The qualifier \enquote{approximately} is important because, strictly speaking, a change in topology is impossible in ideal MHD, where field lines cannot break and reconnect. In astrophysical reality, dissipative processes permit reconnection.

Figures \ref{fig:Topology_Delta01_tfin} and \ref{fig:Topology_Delta02_tfin} show the meridional field lines evolving with time for two choices of $\Delta\Omega$ and $B_0$. At $t = 0.56 t_{\mu}$ (Figure \ref{fig:Topology_Delta01_tfin}) and $t = 0.32t_{\mu}$ (Figure \ref{fig:Topology_Delta02_tfin}), two closed meridional loops appear just above and below the equator. They persist until $t = 1.4 t_{\mu}$ and $t = 1.6 t_{\mu}$ (end of the simulations). Inside the volume of revolution whose meridional cross-section is enclosed by these loops, the field lines have $|B_{\phi}| \gg |B_r|,|B_{\theta}|$ and are approximately \enquote{closed} in the sense described below. Meridional loops, which resemble peninsulas (i.e. which do not pinch off completely), also develop at high latitudes, to the left of the vertical Stewartson layer which is tangent to the inner sphere at the equator \citep{Peralta_Melatos_2009}. Note that a field line is not closed in three dimensions in general, just because its projection onto the meridional plane is closed. For example, the two small, meridional loops near the equator in Figures \ref{fig:Topology_Delta01_tfin} and \ref{fig:Topology_Delta02_tfin} are projections of predominantly toroidal field lines which perform multiple revolutions without closing, as described below. Moreover the meridional loops themselves appear to change their connectedness when the density of contours in the plot increases or the grid resolution in the simulations increases. These small-scale numerical effects are accompanied by small-scale, non-ideal-MHD diffusive effects due to resistivity and viscosity in realistic physical flows. Viscosity and boundary conditions on the surface of these regions control the velocity gradients inside, as opposed to magnetic tension elsewhere, as discussed in Section \ref{sec:integrals}.

In what sense are the equatorial flux tubes enclosed by the loops to be regarded as approximately closed? Inside the closed flux tube, the toroidal magnetic field component typically \mbox{dominates} the poloidal one. We parametrize the magnetic field lines by $[r(\chi),\theta(\chi),\phi(\chi)]$, where $\chi$ is the arc length in magnetic coordinates. For $0 \leq \phi(\chi) \leq 2\pi$, the functions $r(\chi)$ and $\theta(\chi)$ oscillate periodically in $\chi$, with the oscillation period $\chi_{\max}$.
In order to establish whether the magnetic field lines in the flux tube are closed or not, we compare the values of $r(\chi), \theta(\chi)$ and $\phi(\chi)$ at $\chi = \chi_{\min} = 0$ and $\chi = \chi_{\max}$. We consider the magnetic field line to be approximately closed if $|r(0) - r(\chi_{\max})|/r(0) \leq 0.5\%$, $|\theta(0) - \theta(\chi_{\max})|/\theta(0) \leq 0.1\%$ and the field line completes at least one orbit [$\phi(\chi_{\max}) \geq 2\pi$]. 

 Table \ref{tab:tab_A} presents examples of three categories of field lines. The field lines termed \enquote{open} connect the inner sphere to itself at different latitudes. Field lines termed \enquote{periodic} are open, predominatly toroidal and orbit around the inner sphere, with $r(\chi), \theta({\chi})$ oscillating periodically as $|\phi(\chi)|$ increases. In this case, $\chi_{\max}$ corresponds to the oscillation period of $r(\chi)$ and $\theta({\chi})$. \enquote{Closed} field lines return approximately to where they started with $|r(0) - r(\chi_{\max})|/r(0) \leq 0.5\%$, $|\theta(0) - \theta(\chi_{\max})|/\theta(0) \leq 0.1\%$ and $\phi(\chi_{\max}) \geq 2\pi$.

\begin{figure}
\begin{center}
\includegraphics[width=8.2 cm, height = 8.2 cm]{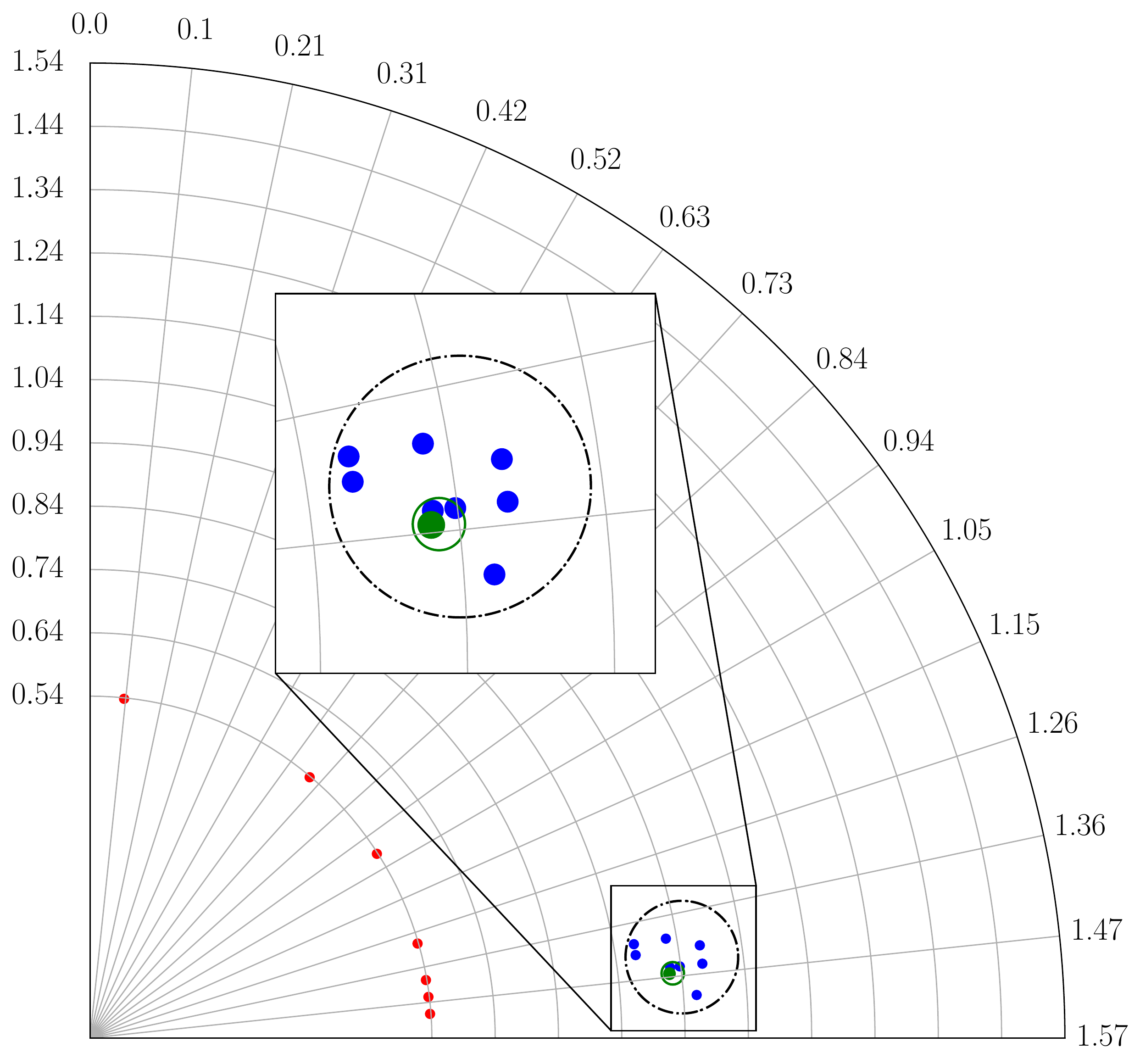}
\caption{Geometry of the magnetic field at $t = 1.4 t_{\mu}$ when the rotational shear is $\Delta\Omega = 0.1$ and the initial configuration is a central point dipole of the form \eqref{eq:dipole_internal_sources} with $B_0 = 1.5\times10^{-2}$. The red dots are the footpoints of open field lines connecting the inner and the outer spheres for $\theta \leq 0.7$ rad. For  $\theta \geq 1.0$ rad the magnetic field lines start at the inner sphere and do not touch the outer sphere. The blue dots represent intersection points with the half-plane $\phi = 0$ for open field lines that do not touch the boundaries and that are almost toroidal. The green dot indicates a closed toroidal field line. The blue dots falling within the green circle drift towards the green dot within ten oscillation periods.}
\label{fig:topology_delta01}
\end{center}
\end{figure}

\begin{figure}
\begin{center}
\includegraphics[width=8.2 cm, height = 8.2 cm]{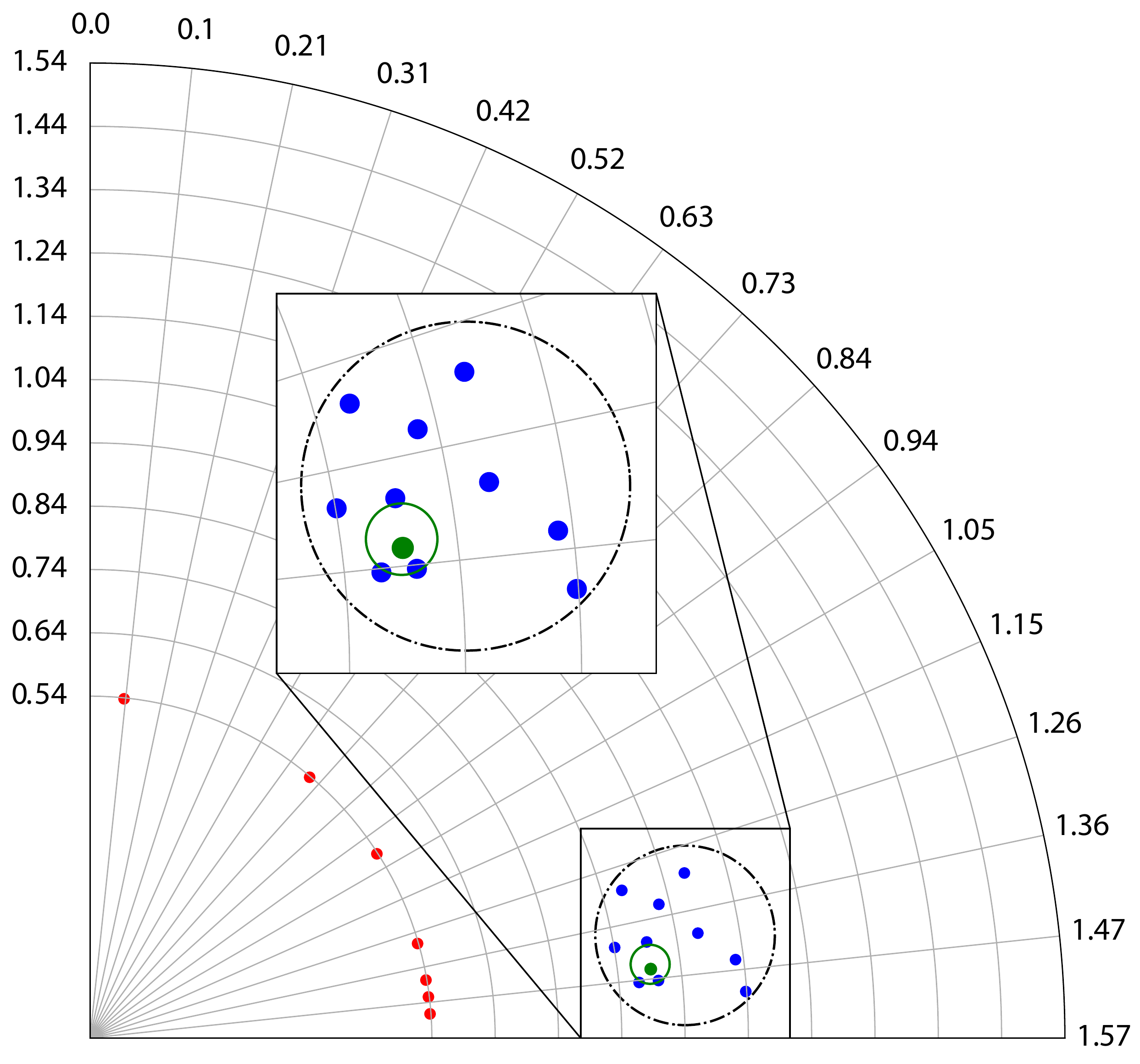}
\caption{As in Figure \ref{fig:topology_delta01} but for $t = 1.6 t_{\mu}$, $\Delta\Omega = 0.2$ and $B_0 = 1.3\times10^{-2}$.}
\label{fig:topology_delta02}
\end{center}
\end{figure}

In Figures \ref{fig:topology_delta01} and \ref{fig:topology_delta02} we plot the coordinates of intersection at $\phi = 0$ of the approximately closed field lines (green dots), open field lines that are predominantly toroidal and touch neither $r = R_i$ nor $r = R_o$ (blue dots), and open poloidal field lines (red dots) as defined in the previous paragraph. In both figures we find that at least one closed toroidal field line forms from the initial open topology \eqref{eq:dipole_internal_sources}. Around the closed line, there is a bundle of predominantly toroidal ($|B_r|,|B_{\theta}| \ll |B_\phi|$) but open field lines, for which $r(\chi)$ and $\theta(\chi)$ oscillate periodically. The oscillation centre stays within the black dot-dashed circle over many periods. The blue dots that fall within the solid green circle containing the green dot deserve a special mention. They also indicate open and periodic field lines, whose barycenter tends towards the closed field line  [over many oscillation periods ($\approx 10$)]. The solid green circle and dot-dashed black circles are smaller for $\Delta\Omega = 0.1$ than for $\Delta\Omega = 0.2$; the higher shear produces a fatter toroidal flux tube. A detailed study of the equatorial flux tubes is presented in Appenidx \ref{sec:topology_flux_tube}.
\begin{table*}
	\caption{Examples of field line endpoints for $B_0 = 1.5\times10^{-2}$, $t = 1.4t_{\mu}$, $\Delta\Omega = 0.1$ (top half) and  $B_0 = 1.3\times10^{-2}$, $t = 1.6t_{\mu}$, $\Delta\Omega = 0.2$ (bottom half) starting from \eqref{eq:dipole_internal_sources} at $t = 0$. For open field lines, $\chi_{\max}$ indicates the maximum value of the arc length at which $r(\chi_{\min} = 0) \approx r(\chi_{\max})$. For periodic field lines $\chi_{\max}$ is the oscillation period of $r(\chi)$ and $\theta(\chi)$, i.e. $r(\chi_{\min}) \approx r$($\chi_{\max}$) and $\theta(\chi_{\min}) \approx \theta$($\chi_{\max}$). For the closed field lines, $\chi_{\max}$ is the arc length at which the field line closes after one or more revolutions. Units: $\chi$ and $r$ are expressed in units of $L$, $\theta$ and $\phi$ are expressed in radians. }
	\label{tab:tableA}
       
	\begin{tabular}{|p{1.5cm}| |p{1.5cm}| |p{1.5cm}||p{1.5cm}|p{1.5cm}|p{1.5cm}|p{1.5cm}|p{1.8cm}}
	\multicolumn{8}{|c|}{}\\
                \hline
		&&&&$\Delta\Omega = 0.1$\\   
		\hline
		 Topology & $\chi_{\max}$ & $r(\chi_{\min})$ & $r(\chi_{\max})$ & $\theta(\chi_{\min})$ & $\theta(\chi_{\max})$ & $\phi(\chi_{\min})$ & $\phi(\chi_{\max}) $\\
		  \hline
		 open  &  0.18872  &     $0.53846 $  &  $0.53863$ & $1.4000$ & $1.7407$ & $0.0$ & $ 0.00018731$\\
                 periodic  &5.4749 &     $0.96069 $  &  $0.96066 $ & $1.5000$ & $1.4991$ & $0.0$ & $-5.9521$ \\ 
                 closed    &11.385 &     $0.92090 $  &  $0.91707 $ & $1.4601$ & $1.4593$ & $0.0$ & $-12.597$ \\  
                 
               \hline
		&&&&$\Delta\Omega = 0.2$\\
		\hline
              open  &  0.076512 &     $0.53846 $  &  $0.53847 $ & $1.5000$ & $1.6416$ & $0.0$ & $-1.4285\times10^{-5}$ \\
               periodic  &  7.8865 &     $1.0272 $  &  $1.0271 $ & $1.4500$ & $1.4546$ & $0.0$ & $-8.04714$  \\
              closed    &  5.4358 &     $0.89243 $  &  $0.89241 $ & $1.4485$ & $1.4485$ & $0.0$ & $-6.3060$ \\
              \hline
	\end{tabular}
	\label{tab:tab_A}
\end{table*}

\subsection{Field-line-averaged-tension}
\label{sec:integrals}
The approximately closed equatorial flux tubes identified in Section \ref{sec:Generating_flux_tube} are not special in the sense that they are out of the ordinary. But they are important because, inside them, the differential rotation is determined differently qualitatively to the rest of the flow. That is, viscosity and boundary conditions on the surface of these regions control the velocity gradients inside, as opposed to magnetic tension elsewhere.

One way to cultivate intuition about the dynamics comes from the field line integrals in Section \ref{sec:line_integrals}. Using the numerical output from PLUTO,
we integrate the left-hand side of equation (\ref{eq:MHD_magn}) along the open and closed magnetic field lines identified in Section \ref{sec:Generating_flux_tube} to obtain

\begin{equation}
I_B = \int{d\chi\gamma(\chi)\vec{B}_p[r(\chi),\theta(\chi)] \cdot \nabla B_{\phi}[r(\chi),\theta(\chi)]}  \ ,
\end{equation}
with

\begin{equation}
\gamma(\chi) = \sqrt{\Big(\frac{d r}{ d\chi}\Big)^2 + r(\chi)^2\Big(\frac{d \theta}{ d\chi}\Big)^2+  r(\chi)^2\sin^2\theta(\chi)\Big(\frac{d \phi}{ d\chi}\Big)^2} \ .
\end{equation}
Along closed field lines with $I_B = 0$, the field-line-averaged momentum equation reads

\begin{equation}
\langle{\partial_tv_{\phi}}\rangle  + \langle\vec{v}\cdot\nabla v_{\phi}\rangle =   \langle\mu \nabla^2 v_{\phi}\rangle
\end{equation} 
and involves only viscous and inertial forces. Along open field lines, e.g. when the initial configuration of the magnetic field is a point dipole of the form \eqref{eq:dipole_internal_sources}, we have $I_B \neq 0$.
 
\begin{figure}
\begin{center}
\includegraphics[width=8.5 cm, height = 5.5 cm]{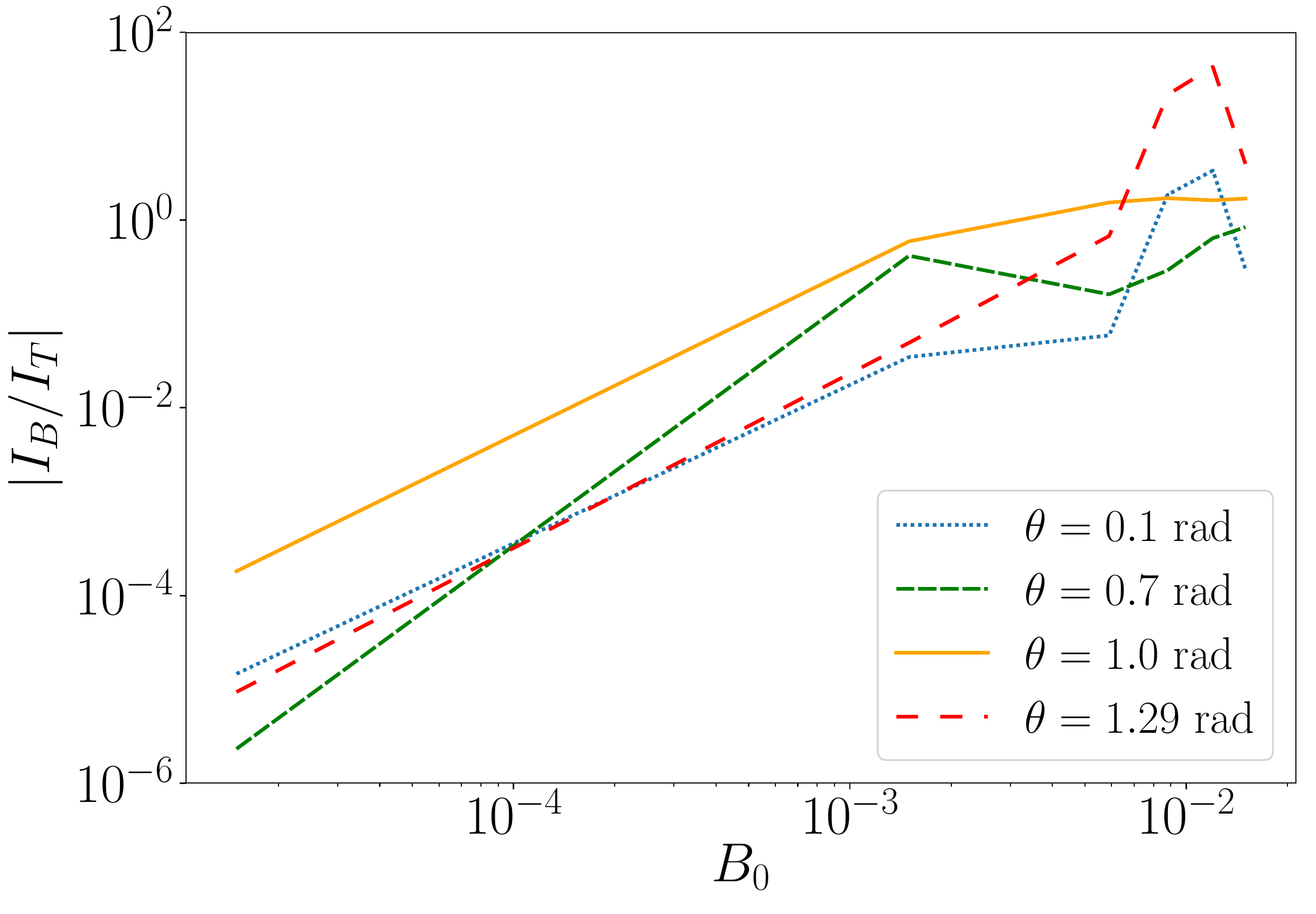}
\caption{Absolute value of $I_B/I_T$ along open field lines with footpoints at $r = R_i$ and $0.1 \leq \theta \leq 1.3$ rad, versus $B_0$ at $t = 1.4 t_{\mu}$ with $\Delta\Omega = 0.1$. The initial configuration is given by \eqref{eq:dipole_internal_sources}. When the ratio reaches unity, magnetic forces are comparable to viscous and inertial forces in a field-line-averaged sense.}
\label{fig:integral_open_delta01}
\end{center}
\end{figure}

\begin{figure}
\begin{center}
\includegraphics[width=8.5 cm, height = 5.5 cm]{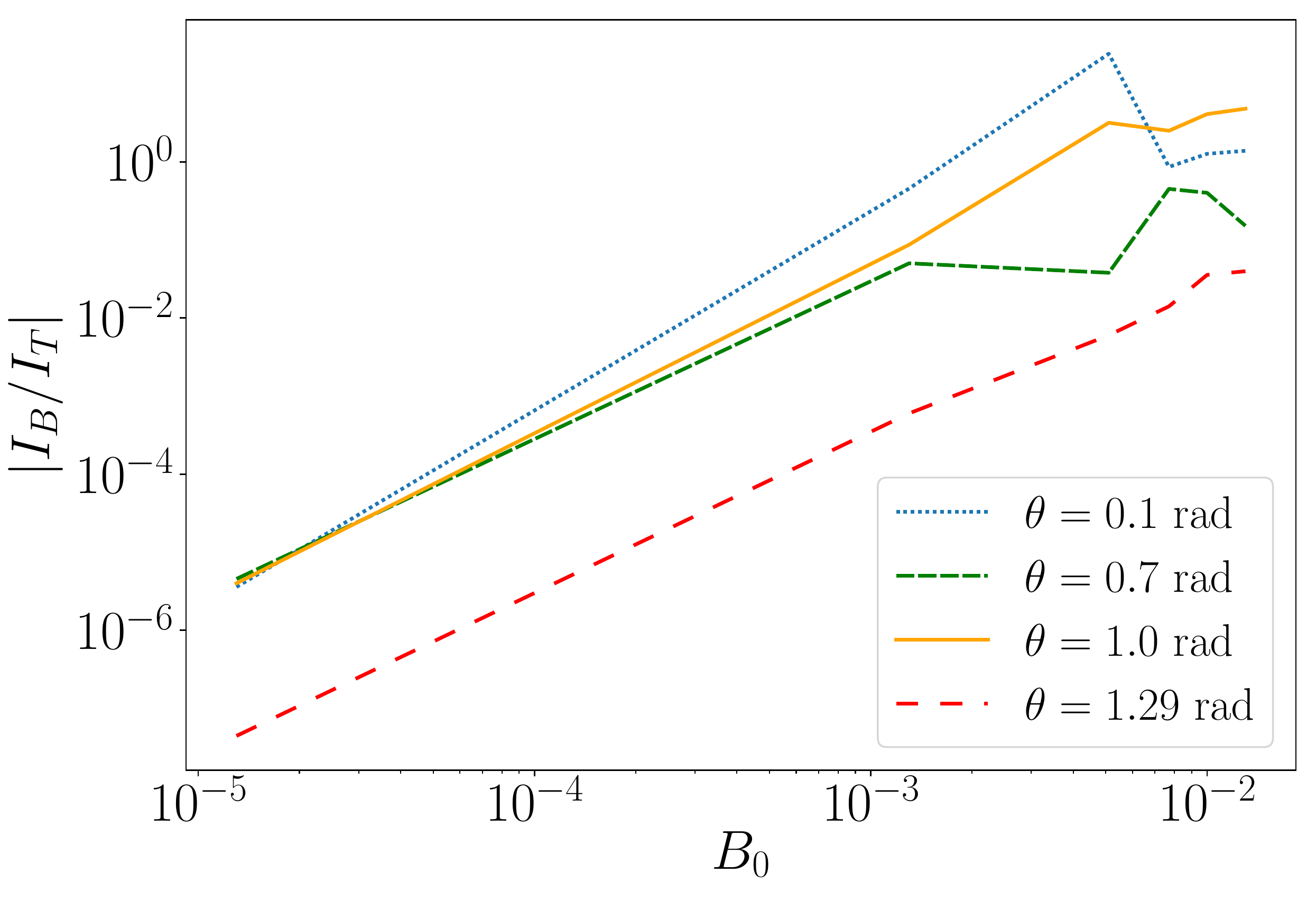}
\caption{As in Figure \ref{fig:integral_open_delta01} with $\Delta\Omega = 0.2$ at $t = 1.6 t_{\mu}$.}
\label{fig:integral_open_delta02}
\end{center}
\end{figure}

 We report in Figure \ref{fig:integral_open_delta01} the value of $I_B$ normalized to

\begin{equation}
I_T = \int{d\chi\gamma(\chi)\bigg[\partial_tv_{\phi}  +(\vec{v}\cdot\nabla)v_{\phi} }  - \frac{1}{4\pi\rho}\vec{B}_p \cdot \nabla B_{\phi} - \mu \nabla^2 v_{\phi}\bigg] 
\end{equation}
as a function of $B_0$ for $\Delta\Omega = 0.1$ and $t = 1.4t_{\mu}$, starting from the point dipole \eqref{eq:dipole_internal_sources}. The ratio $I_B/I_T$ tends to unity as $B_0$ increases and the magnetic coupling dominates the inertial and viscous forces. In Figure \ref{fig:integral_open_delta02} we repeat the same exercise for $\Delta\Omega = 0.2$, starting from the same magnetic footpoints on the boundary. Again, $I_B/I_T$ grows with $B_0$, but $I_B/I_T$ is generally lower than in Figure \ref{fig:integral_open_delta01}. The more the magnetic field tangles, the lower is the line-averaged magnetic tension.

We now evaluate the ratio $I_B/I_T$ for field lines in the closed flux tube. Our choice of the criteria $|r(0) - r(\chi_{\max})|/r(0) \leq 0.5\%$, $|\theta(0) - \theta(\chi_{\max})|/\theta(0) \leq 0.1\%$ to distinguish between purely toroidal, closed field lines and periodic, predominantly toroidal field lines is consistent with $I_B/I_T$. For $\Delta\Omega = 0.1$ (Figure \ref{fig:topology_delta01}), we obtain $I_B/I_T = O(10^{-2})$ for periodic field lines (blue dots) and $I_B/I_T = O(10^{-4})$ for the closed field line (green dot). By doubling the shear (Figure \ref{fig:topology_delta02}), the ratio $I_B/I_T$ reaches $O(10^{-3})$ for periodic field lines and $I_B/I_T = O(10^{-6})$ for the closed field line.

Field line integrals reveal the internal structure of magnetic flux tubes. By inspecting the order of magnitude of the ratio $I_B/I_T$ within the flux tube, it is possible to (i) explore the internal structure of the flux tube and (ii) verify that the field-line-averaged magnetic tension inside the tube is small. Under condition (ii), the angular velocity of the fluid enclosed in the flux tube is determined by the boundary conditions on the surface of the flux tube, which are set by the surrounding fluid. The matching to the flux tube surface occurs via viscous forces on a time scale longer than $t_A$.

\subsection{Nonaxisymmetry}
\label{sec:Nonaxisymmetry}
Nonaxisymmetry arises naturally even in unmagnetized \mbox{spherical} Couette flow [e.g. see the phase plane in Figure $1$ in \citet{Nakabayashi_3} among many other results], and in superfluid spherical Couette flow \citep{Melatos_2007}.

In our paper, we restrict the simulations to relatively low Reynolds number ($Re<10^3$) and to parallel rotation and magnetic axes. We do this partly for simplicity in response to computational constraints and partly because our main application (neutron stars) is arguably a low-Reynolds-number problem with $10^{-10}\leq \Delta\Omega/\Omega\leq 10^{-5}$ from measurements of rotational glitches \citep{Espinoza_1}. However, the low-$Re$ assumption is certainly debatable. One can also argue that the problem is high-Re, because the poorly known effective viscosity of the neutron superfluid may be lower than expected, and glitches may only relax a small fraction of the underlying shear. In that case the flow may be fully turbulent \citep{Melatos_2007}, and a different style of study would be needed.

Assuming low Re and parallel axes, the solution is axisymmetric, and the problem is effectively bidimensional. To test this assertion, we have performed three-dimensional runs, and verified that the solution is independent of the azimuthal coordinate $\phi$. We have performed test runs with $180$ points in the radial direction, $150$ points in the $\theta$ direction and $20$ points in the $\phi$ direction for the configuration $\Delta\Omega = 0.1$ and $B_0 = 1.5\times 10^{-2}$. Both the radial velocity and the azimuthal velocity (not plotted here) are independent of the $\phi$ variable at $r = 1.04$ and $\theta = 0.61, 0.81, 1.21$ rad. So are all the other dependent variables, e.g. magnetic field components. The system starts axisymmetric and remains so as it evolves.

\begin{figure}
\begin{center}
\includegraphics[width=8.5 cm, height = 5.5 cm]{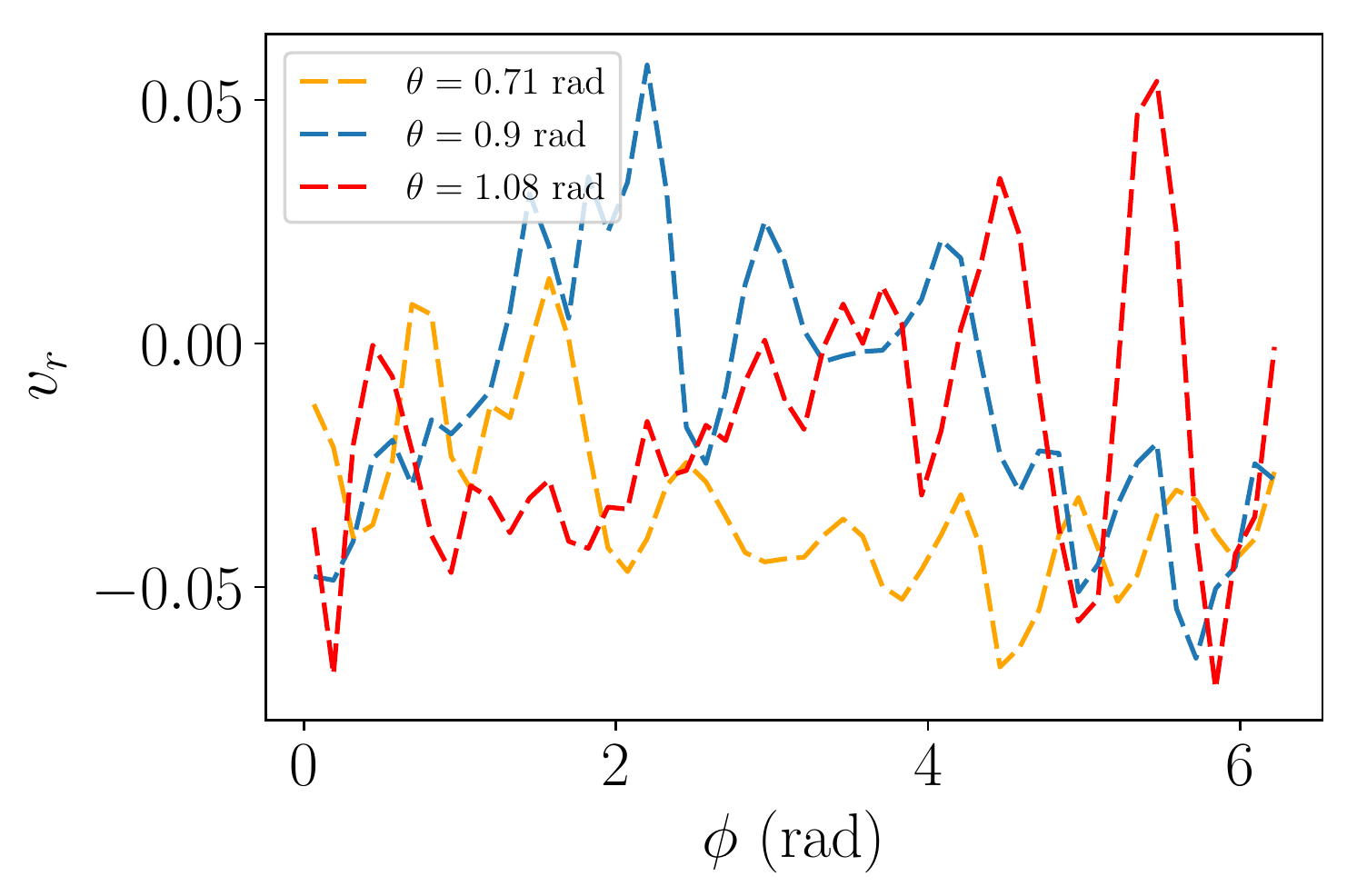}
\caption{Radial velocity as a function of $\phi$ for $\alpha = 0.3$ rad, $\Delta\Omega = 0.1$, $B_0 = 1.5\times 10^{-2}$, $r = 0.99$ and $t = 7.2 t_{E}$, at three latitudes (see legend).}
\label{fig:non_axisymmetry_vr}
\end{center}
\end{figure}

There are two astrophysically realistic ways to drive non-axisymmetry in the flow pattern and magnetic geometry. One way would be to tilt the magnetic axis with respect to the rotational axis, as implied by the radio polarization swings observed in pulsars. The numerical complexity of the problem depends strongly on $B_0, Re,$ the tilt angle $\alpha$, and the grid resolution in $\theta$ and $\phi$ required to resolve turbulent meridional flows and non-axisymmetries. We perform some three-dimensional test runs with resolution $202\times 176\times50$ (in $r$, $\theta$ and $\phi$ respectively) to investigate the behavior informally; a full study lies outside the scope of the paper. We report in Figure \ref{fig:non_axisymmetry_vr} the radial velocity at $\theta = 0.71, 0.90, 1.08$ rad  for $Re = 500$, $\alpha = 0.3$ rad, $\Delta\Omega = 0.1$, $B_0 = 1.5\times 10^{-2}$, $r = 0.99$ and $t = 7.2 t_{E}$, where $t_{E}$ is the Ekman time-scale. There is a clear dependence on $\phi$ at all three latitudes as well as hints of turbulence. In order to resolve the turbulent flow properly, a fine grid is required in $r$, $\theta$ and $\phi$, beyond the capacity of the computational resources at our disposal.

A second astrophysically plausible way to obtain non-axisymmetric flows is to increase the Reynolds number \citep{giss,Hollerbach_2}. Already for Re $ \approx 2000$, non-axisymmetric features emerge even in unmagnetized spherical Couette flow \citep{Nakabayashi_1} and superfluid spherical Couette flow \citep{Peralta_2009}. Again, finer grid resolution is required to resolve the flow. Once one reaches $Re \approx 10^5$, the flow is fully turbulent \citep{Nakabayashi_3, Peralta_2009}. In an ordinary neutron star, where hydrodynamic forces dominate the Lorentz force typically, the magnetic field becomes tangled too, once the flow is fully turbulent. The turbulent fate of the flow is clear qualitatively in this scenario, but several interesting points of detail arise, whose study we postpone for future work, when we gain access to more computational resources.

\section{Conclusion}

An important open question in neutron star astrophysics is how the topology of the star's internal magnetic field affects the degree of differential rotation in the interior.
This paper explores numerically an idealized model of a neutron star consisting of two differentially rotating, concentric spheres containing a magnetized, incompressible fluid described by ideal MHD using the solver PLUTO \citep{Mignone}. Previous \mbox{analytic} studies suggest that the fluid is forced into corotation with the rigid crust, if the field is open, but corotation is harder to maintain everywhere, if the field is closed \citep{Easson1979, article, Lasky}. Differential rotation between the multiple fluid components and the rigid crust is a possible trigger for observed rotational irregularities like glitches \citep{vortex,vortex2,vortex3}.
Numerical investigations of magnetized, spherical Couette flow have been carried out previously in the limit of small magnetic Reynolds number, where the magnetic perturbations induced by the flow are small compared to the externally imposed magnetic field \citep{Dormy, Dormy_Jault, Hollerbach_1,Soward,giss}. In this paper we consider the opposite regime.

When the topology of the field is open and most of the internal fluid is threaded by field lines that connect the northern and southern hemispheres of the outer sphere (i.e. rigid crust), the fluid corotates with the outer sphere after a few Alfv\'en time-scales $t_A$. When open magnetic field lines connect the differentially rotating, inner and outer spheres, some of the fluid at intermediate radii rotates faster than both boundaries. If the inner sphere corresponds \mbox{astrophysically} to a solid inner core \citep{Alford_1, Alford_2, Mannarelli}, fluid along magnetic field lines connecting the inner and outer spheres rotates differentially, as electromagnetic and viscous torques maintain the crust-core rotational lag. If neutron stars do not have solid cores, regions in the inner sphere that are connected magnetically to the crust, approach corotation up to corrections of order $\dot{\Omega}_ot_A$.

It is shown that a central point dipole evolves to generate approximately closed toroidal flux tubes just above and below the equator. The equatorial flux tubes divide into a core (containing at least one approximately closed field line and open, predominantly toroidal and periodic field lines that asymptote to the closed field line) and an annular sheath surrounding it, filled with periodic field lines. For lower rotational shear, the periodic field lines stay within the sheath and out of the core. For higher rotational shear, some field lines leave the sheath, enter the core, and asymptote to the closed field line. The approximately closed equatorial flux tubes develop on roughly the viscous time-scale $t_{\mu}$ and coincide roughly with closed loops in the magnetic field projected into the meridional plane. Inside the approximately closed flux tubes, the magnetic tension integrated along magnetic field lines averages approximately to zero, and the field-line-averaged MHD momentum equation reduces to its hydrodynamical counterpart. Consequently, fluid within the closed flux tube corotates with the surrounding fluid on the viscous time-scale $t_\mu \gg t_{A}$. The magnetic tension does not average to zero along open field lines over a range of $\Delta\Omega$ and $B_0$ values.

We close with two caveats. First, the results of the \mbox{simulations} depend on the initial \mbox{configurations}; different configurations are certain to evolve differently. The fields \eqref{eq:dipole} and \eqref{eq:dipole_internal_sources} are used widely in the general literature \citep{Nataf_1, Schmitt_1, Dormy, Dormy_Jault, Hollerbach_1, Soward} and neutron star models \citep{Bocquet_1, Braithwaite_1}. They lead to two phenomena, which are potentially important astrophysically: \textit{(i)} in some regions the fluid rotates faster than both the inner and outer spheres; and \textit{(ii)} in some magnetically disconnected regions the angular velocity is determined predominantly by viscous forces instead of magnetic tension, as predicted by others analytically \citep{Easson1979, article, Lasky}. It is too early to say whether or not these \mbox{phenomena} are generic to most initial configurations, but their existence for one plausible configuration is interesting for future neutron star \mbox{modeling}. Alternative scenarios, e.g. fully developed turbulence, require a different sort of study.

The second caveat is that the field evolution presented in this paper has been observed before in previous studies of magnetized spherical Couette flow, albeit usually in the regime where the field component induced by the flow is small compared to the applied field. Nevertheless the results are an instructive addition to the neutron star literature. For a long time now it has been customary for neutron star models to assume (for simplicity) a simple internal magnetic geometry, e.g. uniform magnetization or star-centred dipole. There is nothing wrong with this as a theoretical device of course, but Section \ref{sec:flux_tube} reminds readers that the reality is likely to be more complicated. In particular it illustrates that even relatively small amounts of crust-core differential rotation produce complicated magnetic geometries, a point which is fairly straightforward but nevertheless has not received much attention via numerical simulations in the literature. In these more complicated regions, the physics governing differential rotation is qualitatively different to elsewhere, with viscosity dominating magnetic tension as discussed above. The results in Section \ref{sec:flux_tube} complement other studies on complicated internal magnetic fields, e.g. \cite{Braithwaite_1} and \cite{Braithwaite_2}, who did not consider \mbox{differential} rotation but did consider very interesting topologies; \citet{Pons_3}, \citet{Pons_1} and \citet{Pons_2}, who focused on the crust and simulate the core; \cite{Drummond_1, Drummond_2}, who looked at the related problem of spontaneous emergence of complicated neutron vortex structures in a type II superconductor; \cite{Ruderman_1}, who studied surface multipoles and their tectonic evolution; and \cite{Sur_1}, who studied the turbulent equilibrium of the magnetized stellar interior.


\section*{Acknowledgements}
F. Anzuini acknowledges the support of the University of Melbourne through a Melbourne Research Scholarship. The authors thank the anonymous referee for pointing out several relevant references on magnetized spherical Couette flow and for helpful suggestions on how to restructure and sharpen parts of the presentation. 
\bibliographystyle{mnras}
\bibliography{BIBLIO}


\appendix

\section{Angular velocity profiles}
\label{sec:all_dipole}

\label{sec:all}
\begin{figure*}
\includegraphics[width=14cm, height = 8.0 cm]{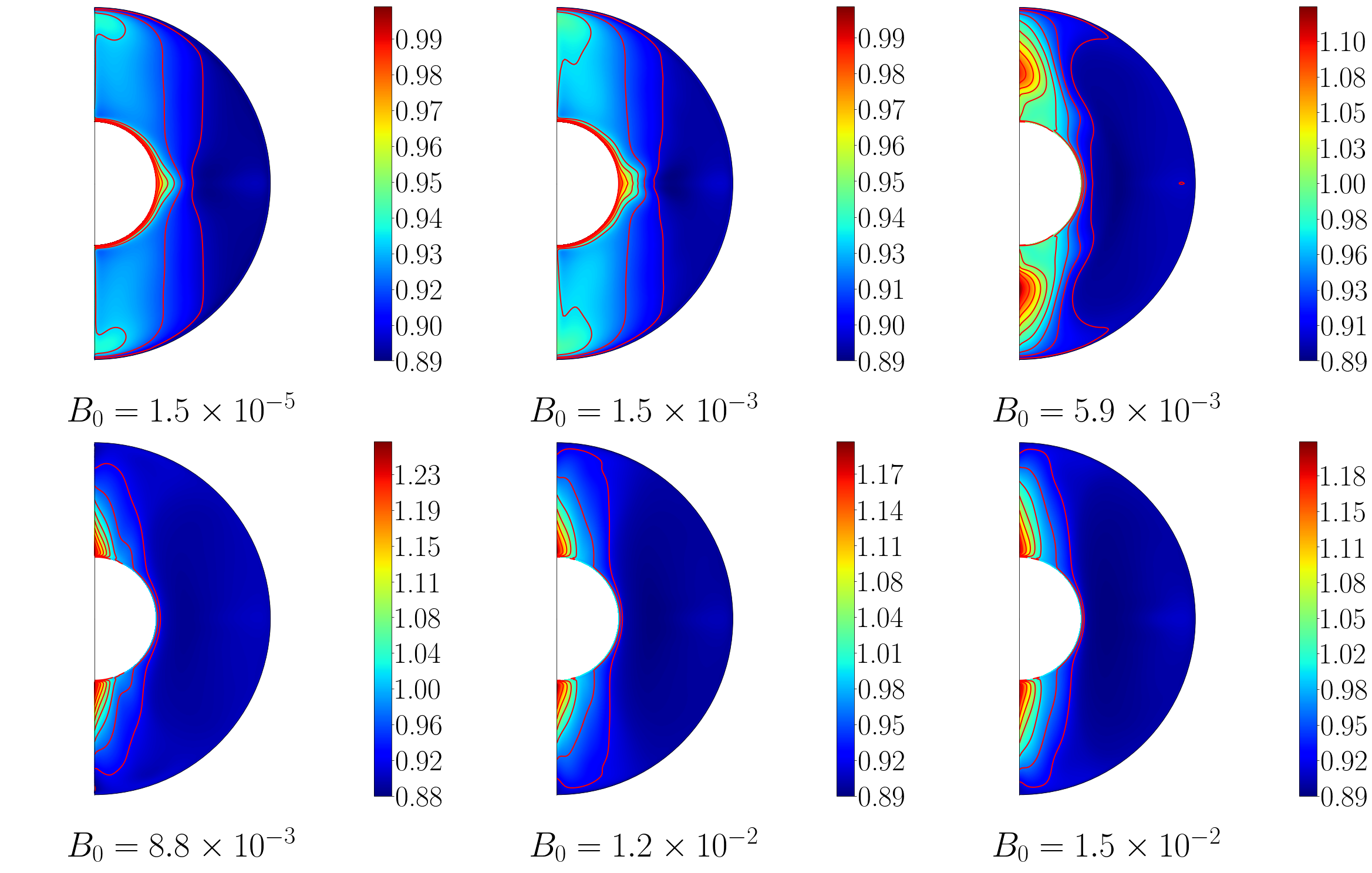}
\caption{Contour plots of the angular velocity of the fluid at $t = 1.4 t_{\mu}$ for $\Delta\Omega = 0.1$ and an initial dipole given by (\ref{eq:dipole}). Contours of $\Omega$ are represented as red solid curves. As the initial magnetic field strength rises, we observe $\max(\Omega) > \Omega_i$ near the poles. For $ 0.4 \lesssim \theta \lesssim 5.9$ rad, the fluid corotates approximately with the outer sphere.}
\label{fig:omega_all_dipole_delta01}
\end{figure*}

\begin{figure*}
\includegraphics[width=14cm, height = 8.0 cm]{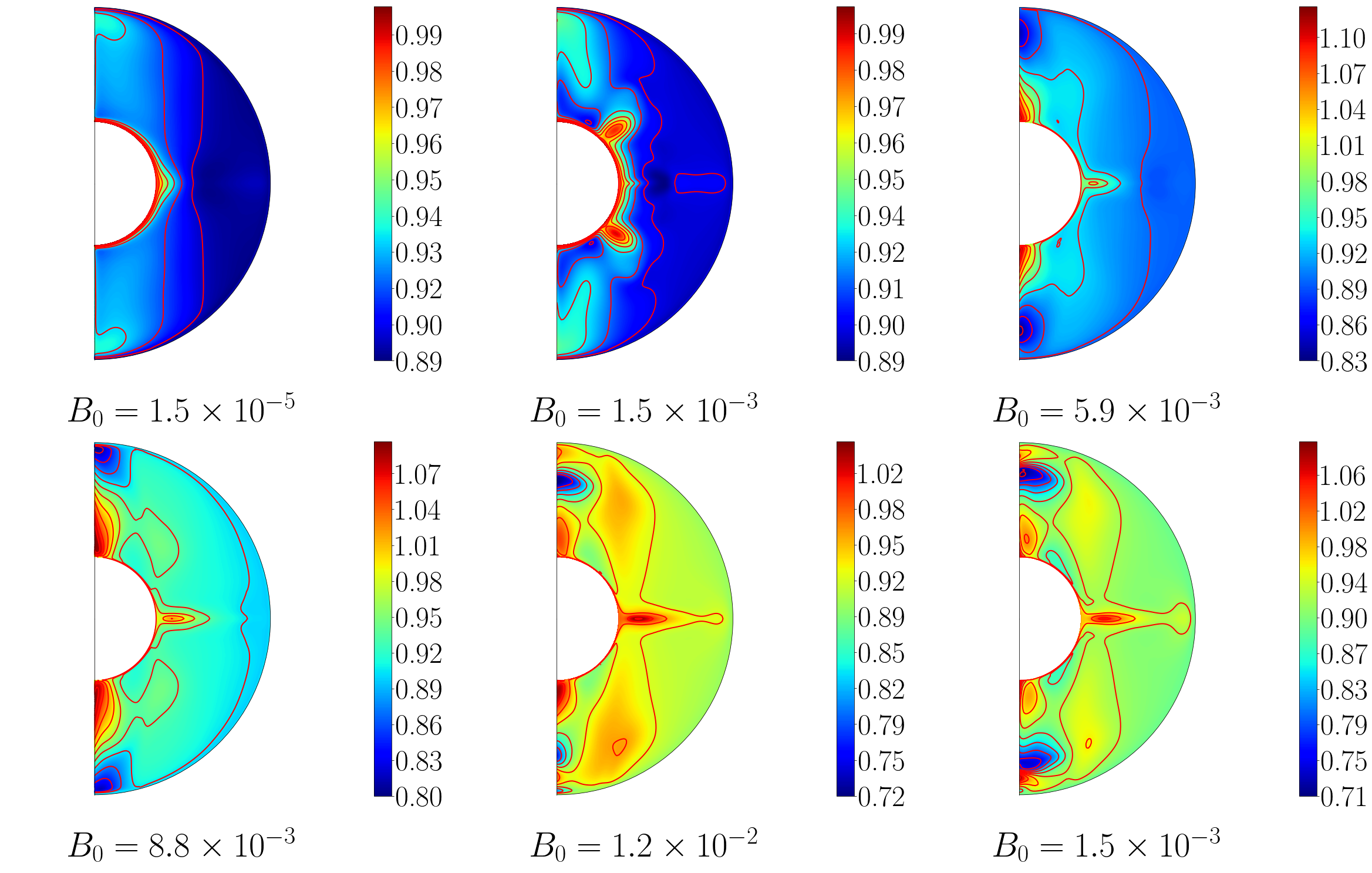}
\caption{Contours of the angular velocity of the fluid (red curves) for $\Delta\Omega = 0.1$ at $t = 1.4 t_{\mu}$. As the initial magnetic field strengthens, $\Omega > \Omega_i $ for $B_0 \geq 5.9\times10^{-3}$. A small north-south asymmetry is observed at the poles for $B_0 \geq 1.2\times10^{-3}$. }
\label{fig:omega_all_delta01}
\end{figure*}

\begin{figure*}
\includegraphics[width=14cm, height = 8.0 cm]{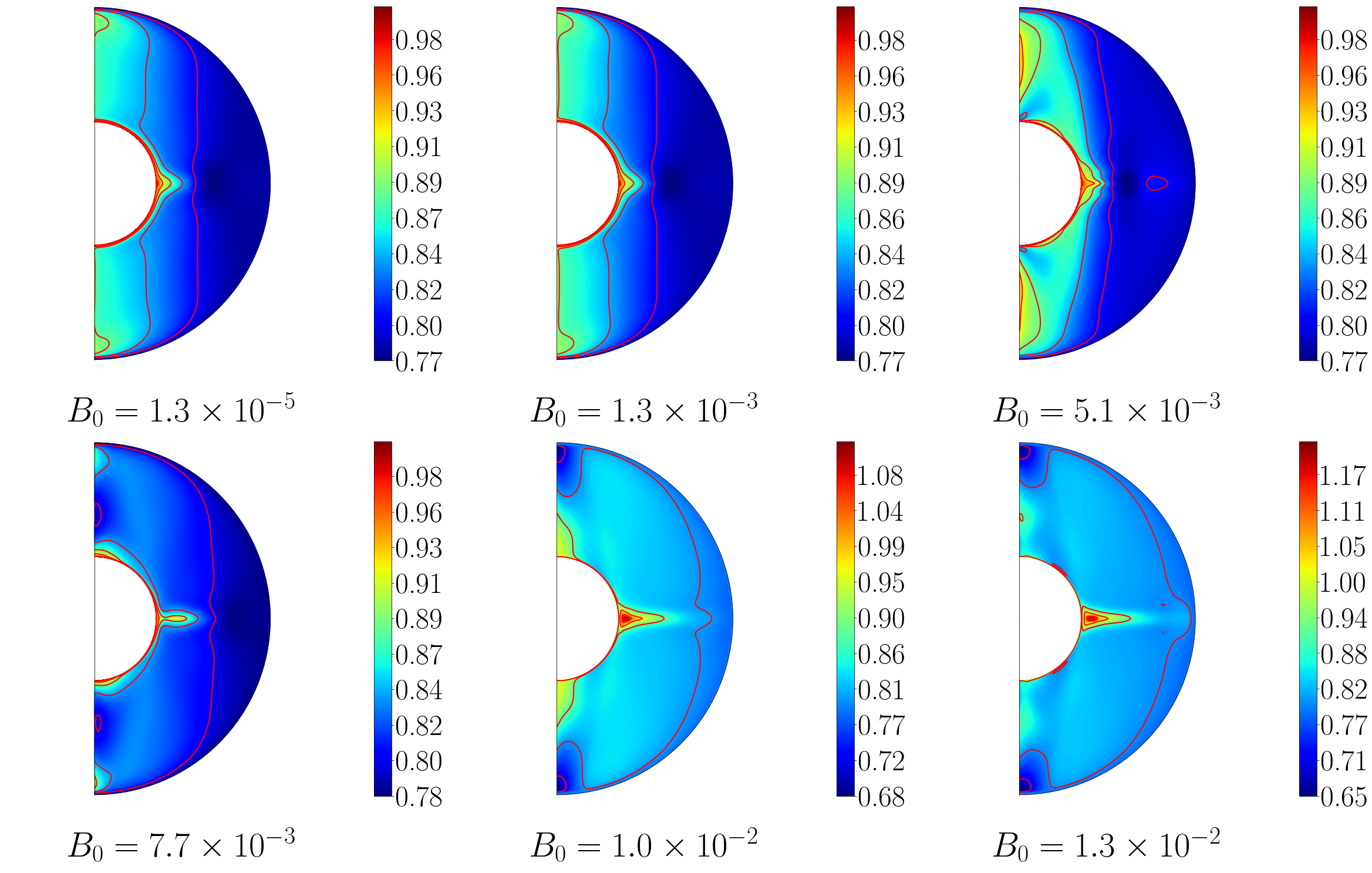}
\caption{As in Figure \ref{fig:omega_all_delta01} but for $\Delta\Omega = 0.2$ at $t = 1.6 t_{\mu}$.}
\label{fig:omega_all_delta02}
\end{figure*}

In this appendix we show for completeness $\Omega(r,\theta)$ for different $B_0$, $\Delta\Omega$ and for the initial magnetic field topologies given by \eqref{eq:dipole} and \eqref{eq:dipole_internal_sources}. In Figures \ref{fig:omega_all_dipole_delta01}, \ref{fig:omega_all_delta01} and \ref{fig:omega_all_delta02} we plot the contours of the angular velocity (color bar in units of $\Omega_i$). The red curves are the contour levels of $\Omega$.

Figure \ref{fig:omega_all_dipole_delta01} corresponds to the initial magnetic field configuration given by \eqref{eq:dipole}. As expected, a weak magnetic field does not lead to appreciable differences between the magnetized and the unmagnetized cases. For $B_0 \geq 1.5\times10^{-3}$, differences are clearly observed. The fluid spins faster than the inner sphere at the poles for $B_0 = 5.9\times10^{-3}$, and $\Omega$ keeps growing up to $B_0 = 8.8\times10^{-3}$. For $ 0.4 \lesssim \theta \lesssim 5.9$ rad, the fluid is approximately in solid body rotation with the outer sphere, independently of $B_0$. The border separating regions with $\max(\Omega) > \Omega_i$ from corotating regions is delimited, for increasing $\theta$, by the first magnetic field line that does not touch the inner sphere. 
The presence of poloidal closed field lines in the initial configuration does not prevent corotation with the outer sphere.

Figures \ref{fig:omega_all_delta01} and \ref{fig:omega_all_delta02} show $\Omega(r, \theta)$ across the spherical shell for $\Delta\Omega = 0.1$ and $\Delta\Omega = 0.2$ and $1.3\times 10^{-5} \leq B_0 \leq 1.5 \times 10^{-2}$.
For $\Delta\Omega = 0.1$ and  $B_0 \approx 10^{-5}$, the solution is indistinguishable from the unmagnetized case in Section \ref{sec:Viscous}. For $B_0 \geq 5.9\times10^{-3}$, we obtain $\Omega > \Omega_i$ in certain regions. For $B_0 \geq 1.2\times10^{-2}$, a small north-south asymmetry is observed, which seems numerical in origin.

In Figure \ref{fig:omega_all_delta02}, for $\Delta\Omega = 0.2$, we find $\max(\Omega) >\Omega_i$ for $B_0 \geq 1.0\times10^{-2}$. Unlike $\Delta\Omega = 0.1$, the locations with $\max(\Omega) > \Omega_i $ lie near the equator rather than at the poles \citep{Soward}.
As for \eqref{eq:dipole}, the fluid rotates differentially along open field lines connecting the inner and outer spheres.

\section{Detailed structure of the approximately closed equatorial flux tubes}

\begin{figure*}
  \begin{minipage}{0.43\textwidth}
  \centering
    \includegraphics[width=8.2 cm, height = 8.3 cm]{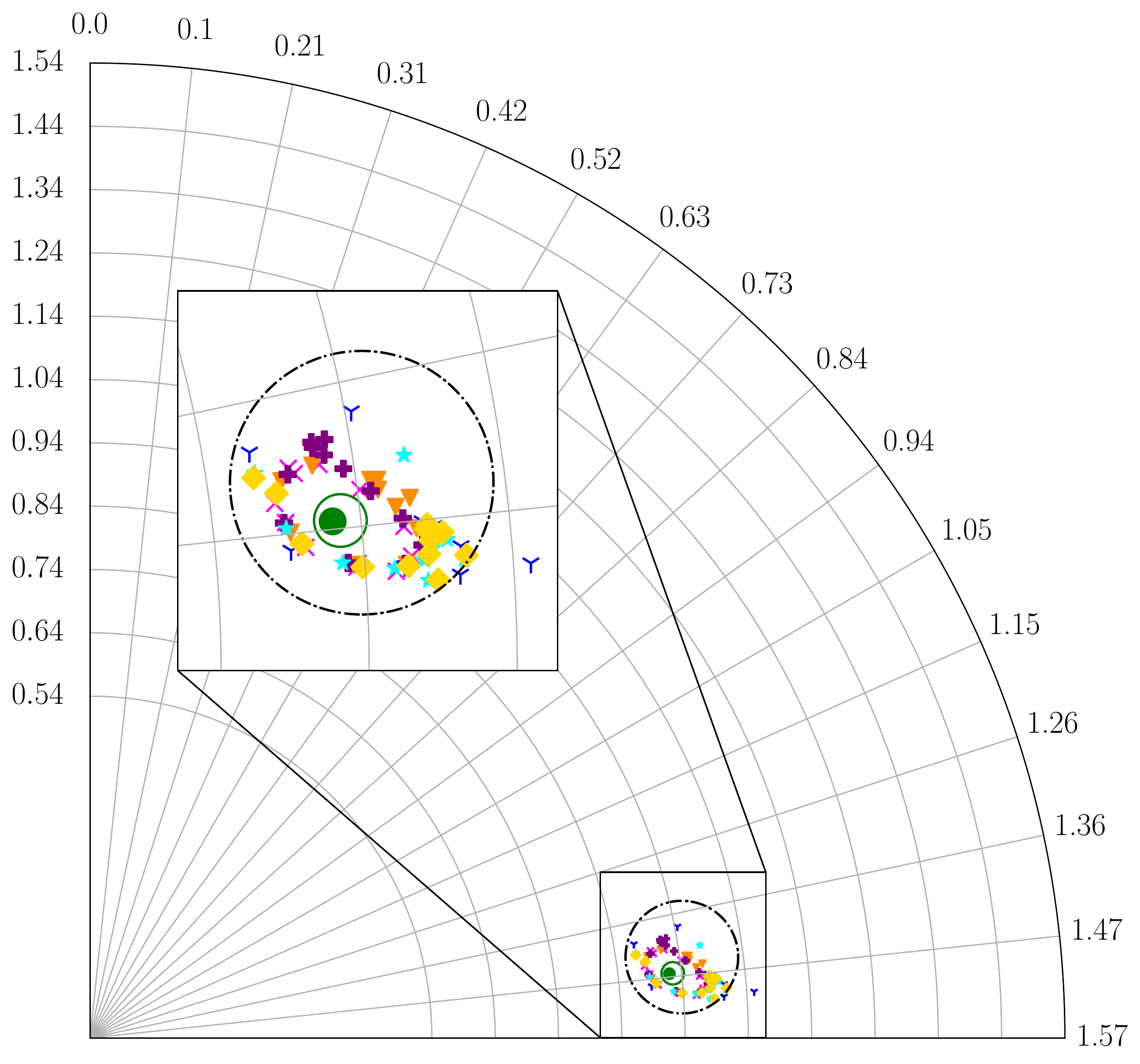}
    \caption{Successive intersection points with the $\phi = 0$ half-plane of the blue dots outside the solid green circle shown in Figure \ref{fig:topology_delta01} for $\Delta\Omega = 0.1$ and $B_0 = 1.5 \times 10^{-2}$. Every intersection point corresponds to a $2\pi$-revolution in $\phi$. There are 12 revolutions shown per field line. Each color marks a separate field line. The black dot-dashed circle is densely filled with points, but no field line strays into the green circle. }
    \label{fig:subsequent_footpoints_delta01}
  \end{minipage} \qquad \qquad
  \begin{minipage}{0.43\textwidth}
  \centering
    \includegraphics[width=8.2 cm, height = 8.3 cm]{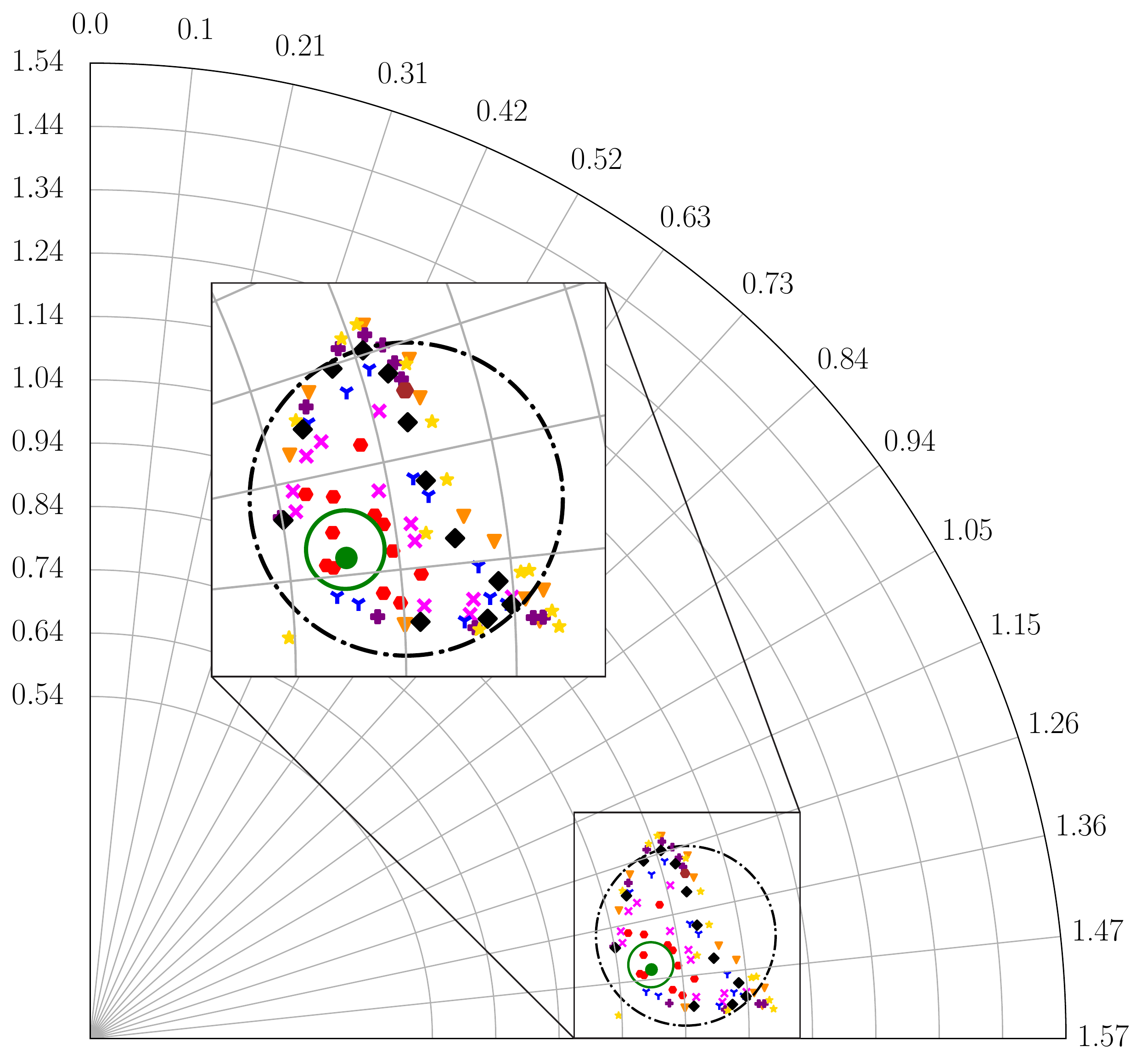}
    \caption{As in Figure \ref{fig:subsequent_footpoints_delta01} but for $\Delta\Omega = 0.2$ and $B_0 = 1.3\times10^{-2}$. Unlike in Figure \ref{fig:subsequent_footpoints_delta01}, some of the intersection points enter the green solid circle. \newline \newline \newline \newline}
    \label{fig:subsequent_footpoints_delta02}
  \end{minipage}
\end{figure*}

In Section \ref{sec:Generating_flux_tube} we find that field lines that lie within the green circles in Figures \ref{fig:topology_delta01} and \ref{fig:topology_delta02} stay within the green circles and indeed asymptote to the nearly closed field lines marked by a green dot. We now ask what happens to field lines that lie outside the green circle but inside the dot-dashed black circle. The question is studied in Figure \ref{fig:subsequent_footpoints_delta01} for $\Delta\Omega = 0.1$. We choose a new color for each blue dot outside the solid green circle in Figure \ref{fig:topology_delta01} and follow 12 field line revolutions for each.

We see in Figure \ref{fig:subsequent_footpoints_delta01} that the successive intersection points with the half-plane $\phi = 0$ stay within the black dot-dashed circle and swirl around the green circle without entering the green circle. In other words, the geometry near the green dot separates into two parts: a core, where field lines asymptote to the green dot, and a separate annular sheath around the core. Similarly, in Figure \ref{fig:subsequent_footpoints_delta02} we track 12 revolutions for each of the blue dots outside the green circle in Figure \ref{fig:topology_delta02} for $\Delta\Omega = 0.2$. The magnetic geometry is richer than for $\Delta\Omega = 0.1$. The magnetic field lines spread across a larger volume, and the clustering of intersection points is less dense. Some points (red hexagons) enter the green circle after starting outside it unlike in Figure \ref{fig:subsequent_footpoints_delta01}, and their barycenter tends subsequently towards the closed field line.

In summary, a magnetic field given by \eqref{eq:dipole_internal_sources} evolves to generate a toroidal equatorial flux tube. Inside the flux tube, we find that one purely toroidal, closed field line is surrounded by a bundle of predominantly toroidal but open field lines. In order to qualify as closed, the field line meets the conditions $|r(0) - r(\chi_{\max})|/r(0) \leq 0.5\%$ and $|\theta(0) - \theta(\chi_{\max})|/\theta(0) \leq 0.1\%$. In Section  \ref{sec:integrals}, we show that these thresholds can be related to field line integrals; if $r(\chi)$ and $\theta(\chi)$ satisfy the above conditions, the field-line-averaged magnetic tension vanishes.





\bsp	
\label{lastpage}
\end{document}